\documentclass[9pt,twocolumn,twoside]{pnas-new}

\usepackage{bm}
\usepackage{caption,subcaption}
\usepackage{adjustbox}

\newcommand{\RM}{{\mathbb R}}

\newcommand{\EM}{{\mathbb E}}

\newcommand{\Cc}{{\mathcal C}}

\templatetype{pnasresearcharticle} 

\title{Quantifying Gibbs measures of disordered crystals up to the solid-liquid phase transition}

\author[a,1]{Vladislav Efremkin}
\author[b,1]{Julian Heske}
\author[a,c,2]{Thomas D. K\"uhne}
\author[d]{Emil Prodan}

\affil[a]{Center for Advanced Systems Understanding, Helmholtz Zentrum Dresden-Rossendorf, D-02826 G\"orlitz, Germany}
\affil[b]{Department of Chemistry, Paderbon University, D-33098 Paderborn, Germany}
\affil[c]{Institute of Artificial Intelligence, TU Dresden, D-01187 Dresden, Germany}
\affil[d]{Department of Physics, Yeshiva University, New York, NY 10016, USA}

\leadauthor{Vladislav Efremkin} 

\significancestatement{We propose and support the conjecture that a crystalline phase of a condensed matter system with a complicated structural phase diagram is fully and uniquely encoded in a distinct subset of facets of the Voronoi cells that are stable against the thermal fluctuations throughout the entire part of the phase diagram corresponding to that crystalline phase.}

\authorcontributions{V.E. performed research, analyzed data and contributed to writing the paper, J.H. performed research and analyzed data, T.D.K. designed research and contributed to performing the research, analyzing the data and writing the paper, E.P. designed research, analyzed data and wrote the paper.}
\authordeclaration{Please declare no competing interests.}
\equalauthors{\textsuperscript{1}V.E. (Author One) contributed equally to this work with J.H. (Author Two).}
\correspondingauthor{\textsuperscript{2}To whom correspondence should be addressed. E-mail: tkuehne@cp2k.org}

\keywords{Gibbs measure $|$ ab-initio molecular dynamics $|$ disordered solids $|$ solid-liquid phase transition $|$ Voronoi tessellation } 

\begin{abstract}
Quantifying the configuration space and the Gibbs measure of thermally disordered condensed matter systems has been a long standing problem. The challenge is to avoid the Gibbs paradox, which forbids any ordering or labeling of the atoms. Our key observation is that the lattice of a thermally disordered condensed matter system, in either solid, liquid or gas phase, can be fully reconstructed from the Voronoi cells of the atoms alone, even if these Voronoi cells are disassembled and randomly scrambled. In the example of the crystalline phase of silicon, the statistics of the Voronoi cells  reveals the existence of four, and only four, large facets that are present with probability one for all temperatures up to the solid-liquid melting line. These four largest facets, which separate nearest-neighboring atoms, can be also be used to reconstruct the lattice of the crystal. Hence, their collection supplies the optimal representation of the configuration of the crystal. We conjecture that the existence of Voronoi facets that, despite their large thermal fluctuations, survive with probability one up to the melting temperature, is the fundamental signature of the crystalline solid phase and therefore key to quantifying the Gibbs measure over the entire solid phase.
\end{abstract}

\dates{This manuscript was compiled on \today}
\doi{\url{www.pnas.org/cgi/doi/10.1073/pnas.XXXXXXXXXX}}

\begin{document}

\maketitle
\thispagestyle{firststyle}
\ifthenelse{\boolean{shortarticle}}{\ifthenelse{\boolean{singlecolumn}}{\abscontentformatted}{\abscontent}}{}


\dropcap{T}he Gibbs measure of a finite physical system with a classical $(\bm p,\bm q)$-phase-space encodes the equilibrium state of the system at a given temperature, quantified here by the usual $\beta$-parameter.\footnote{Given by the inverse of the product between the Boltzmann constant and the actual temperature.} The kinetic contribution to the equilibrium distribution is often integrated out and the reduced Gibbs measure ${\rm d}\mathbb P(\bm q) = Z^{-1} e^{-\beta V(\bm q)} d \bm q$ is used instead, hereafter simply referred to as Gibbs measure. It quantifies the probability of finding the system in different areas of the configuration space, {\it i.e.} within the allowed range of the generalized coordinates $\bm q$. In many settings of interest, such as that of first-principles molecular dynamics (FPMD) \cite{car1985unified, sugino1995ab}, the ergodic component of the Gibbs measure corresponding to a pure phase is conveniently derived from the Langevin dynamics at fixed temperature, via Brikhoff's relation 
\begin{equation}
  \mathbb P(Q) = \lim_{T \to \infty} \frac{1}{T}\int_0^T \chi_Q\big ( {\bm q}(t,\beta) \big ) d t, 
\end{equation} 
where $Q$ is a subset of the configuration space, $\chi_Q$ is its indicator function and $\bm q(t,\beta)$ is a thermalized orbit at inverse temperature $\beta$ (see Materials and Methods).

For condensed phase systems, such as crystalline silicon investigated here, quantifying the Gibbs measure of the classical degrees of freedom is essential for the accurate simulation of macroscopic electronic properties, because the electron dynamics is determined by their thermally disordered atomic configurations. Indeed, as demonstrated in \cite{KP2018,KHP2020}, effects such as Anderson localization of the electronic wave functions can strongly influence the quantitative values of the transport coefficients because the positions of the mobility edges of the electronic spectra are very sensitive to the amount and type of disorder. In principle, their macroscopic electronic properties can be computed on-the-fly during a FPMD simulation, but such simulations are computationally very demanding. 
Having the Gibbs measure quantified at various temperatures will certainly enable a broader scientific community to generate accurate thermalized atomic configurations, hence to simulate the electronic properties of materials in settings that mimic the actual conditions in which electronic devices operate.

However, the Gibbs measure is desirably defined over a {\it compact} configuration space in order to be normalisable, but this is in apparent conflict with the {\it non-compact} character of the crystal's lattice. Ref.~\cite{KP2018} supplied a specific solution to this puzzle that relies on the fact that, at temperatures well below the melting temperature, the temporal orbits of the atoms spend a long time\footnote{Greatly exceeding the temporal length of typical FPMD simulations.} wandering around an average position. As such, the configuration space of the crystal can be modeled as a product of small point particles centered at the origin of $\RM^3$, by rendering the orbits of the atoms from their average centers. Tychonoff's theorem then assures that this space is compact. Unfortunately, this becomes problematic at temperatures close to the melting line, where the atoms start to exchange positions. Nevertheless, the FPMD simulations of Ref.~\cite{KP2018} revealed the striking fact that the Gibbs measures for thermally disordered states of the silicon crystal can be quantified with an unexpectedly high degree of accuracy using no more than just five parameters! This is significant because, before that study, it was believed that the Gibbs measure  derived from FPMD simulations can be represented by long nuclear trajectories only.

In the present work, we undertake the challenge of quantifying the Gibbs measure of the silicon crystal all the way to its melting temperature. The first challenge is removing any label from the indistinguishable atoms.\footnote{This will, of course, also keep us safe from the Gibbs paradox.} Thus, a thermally disordered configuration of the atoms should be seen as nothing more than a discrete subset, hence a pattern in the physical space $\mathbb R^3$. Pattern recognition and pattern analysis is a mature scientific field and many tools can and have been borrowed from there and applied to materials science \cite{Bellissard2015}. For example, the set of compact subsets of $\mathbb R^3$ can be topologized by endowing it with the Hausdorff metric \cite{Barnsely1993}. The class of closed infinite subsets of $\mathbb R^3$, such as the atomic configurations in the thermodynamic limit, can be topologized in a similar way by employing the one-point compactification of $\mathbb R^3$. It is well known that this leads to a compact space of patterns \cite{Lenz2003}, hence to a formal solution to one of our stated challenges. In this setting, we are no longer dealing with the many orbits of the atoms in the physical space, but rather with the (temporal) orbit of one Delone set in this abstract topological compact space of patterns. This point of view has been developed and popularized by Jean Bellissard \cite{Bellissard2015}, who made the deep statement that any homogeneous state of matter can be identified with a probability measure over the space of Delone subsets of $\RM^3$, which is ergodic against the spatial translations. For practical purposes, however, we need an explicit characterization of the configuration space that supports the Gibbs measure and this is the main practical obstacle we or anybody else needs to overcome. 

We present a complete solution to the stated challenges, which engages the Voronoi tessellations of the thermally fluctuating atomic lattices. Working with silicon as an example, we demonstrate that the configuration space of the atoms can be encoded in the geometry of the Voronoi cells and that the Gibbs measure can be quantified over a submanifold of this configuration space. Besides the practical aspects, our technique supplies a fresh insight into the very nature of crystalline phases. Let us recall that the crystalline structure, despite large thermal fluctuations, is detectable in X-ray diffraction patterns for temperatures up to the melting line of a crystal \cite{FittingJOM1999}. While such X-ray diffraction patterns can be used to identify the crystalline phases of condensed matter systems with a complex structural phase diagram, it is also desirable to have real-space fingerprints markers for each of the crystalline phases. For the silicon crystal, we find such markers in the fluctuating geometries of the Voronoi cells, specifically, in the existence of exactly four facets that remain stable for temperatures up to silicon's melting line. We conjecture that such real-space markers exist for any crystalline phase.

\subsection*{Encoding the configuration space using pointed Voronoi cells} 

A Voronoi cell of a homogeneous system is a convex shape bounded by $N$ flat surfaces, with $N$ not exceeding a firm upper bound $\bar N$. A pointed Voronoi cell is a Voronoi cell together with the unique atom enclosed by it. Relative to this reference point, a Voronoi cell can be quantified by a number $N$ of 3-component vectors $\bm v_n$, encoding the positions of the $N$ points of the $N$ facets that are closest to the reference point. Given these points, the Voronoi cell's facets can be reconstructed by drawing the planes passing through these points perpendicularly to their  corresponding $\bm v_n$ vectors. Thus, a pointed Voronoi cell can be quantified using at most $3\bar N$ real numbers.

The next important fact is that the Voronoi cells do not need to carry any special label. Indeed, suppose we take a snapshot of a thermalized atomic configuration of a crystal at some finite temperature and then we decompose the sample into its Voronoi cells, which we place in a bucket. We then scramble the bucket to erase any history on the spatial locations and orientations of the cells. Mathematically, this is equivalent to dealing with an unordered set of Voronoi cells. The key question is: Can we uniquely reconstruct the crystal? The answer is yes! For this, we pick a Voronoi cell from the bucket and we search for a second Voronoi cell with a facet that perfectly matches one of the facets of the first Voronoi cell. We then glue together these two cells along that specific facet and repeat the process. As one can see, gluing cells along matching facets, together with the assurance that the matching pairs of facets are almost surely unique, enables us to reconstruct the whole atomic configuration, a process that works for the crystalline, as well as for the liquid phases.\footnote{We have verified this statement for a large pool of atomic configurations.} This is our first major finding that we want to communicate in this article:

{\bf Statement 1:} {\it The Gibbs measure of the classical degrees of freedom of a material is fully encoded in the statistics of the pointed Voronoi cell geometries.}

Let us point out that the reconstructed atomic lattice from the randomized Voronoi cells will have arbitrary orientation in space, which is fully determined by the arbitrary orientation of the Voronoi cell that initializes the reconstruction process. However, the macroscopic orientation of the lattice is preserved by the time evolution, and each orientation can occur with equal probability.

\subsection*{Unique and complete markers for crystalline phases} 

\begin{figure}
\centering
\includegraphics[width=\linewidth]{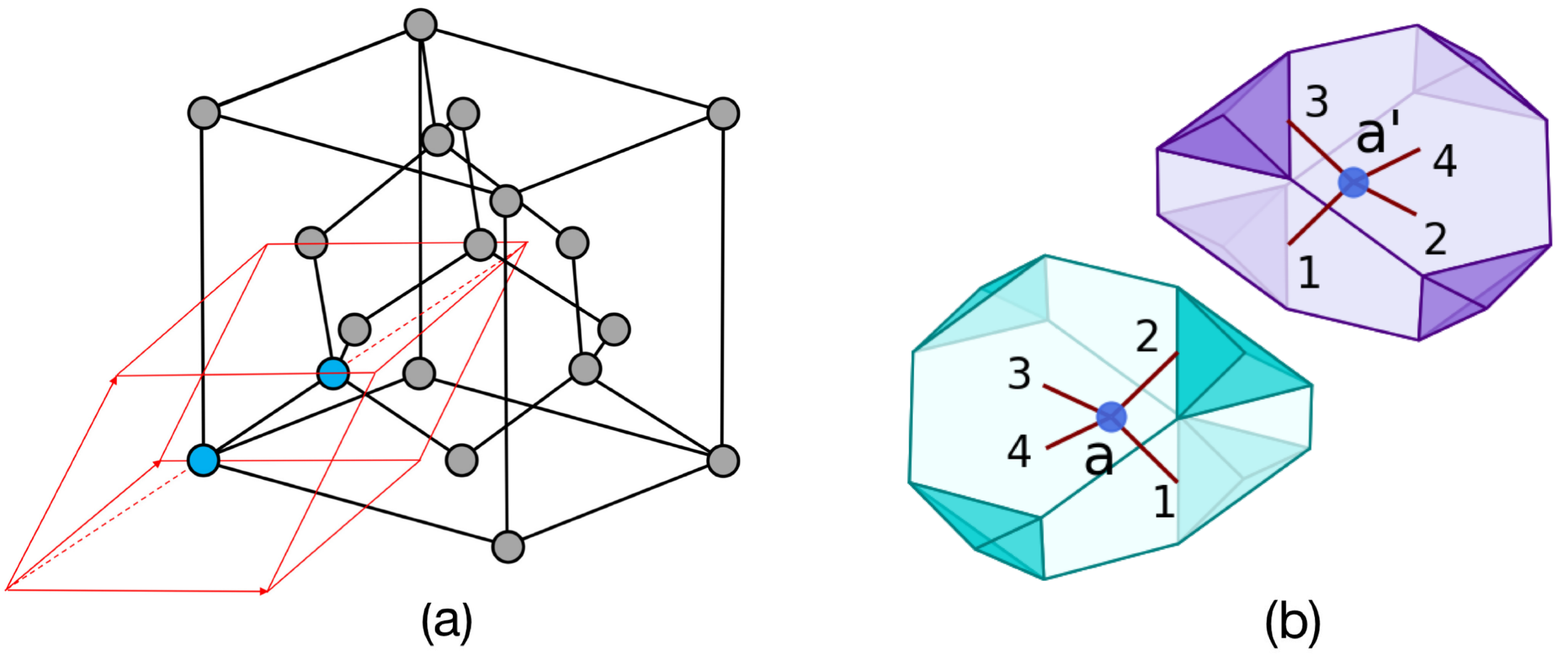}
\caption{Silicon crystal's structure at zero temperature. (a) The unit cell (black cube) and the primitive cell (red rhombohedron); (b) The Voronoi cells of the two atoms in the primitive cell.}
\label{Fig:SiOverview}
\end{figure}

As we shall see, the Voronoi cells of ordinary materials can be unexpectedly complex and further insight is needed to make the problem conceptually and numerically tractable. As a working example, we chose here to showcase silicon in its both crystalline and liquid phases. 
The crystalline structure at $0$K is shown in Fig.~\ref{Fig:SiOverview}, together with the Voronoi cells of the two atoms in the repeating Bravais lattice cells. We recall that a single Voronoi cell at 0K tiles the entire $\RM^3$ space, when acted on by the elements of the space symmetry group $Fd\bar 3 m$ of the silicon crystal \cite{Singh93,wondratschek2004international}. Thus, at finite temperatures, apart from thermal fluctuations, all Voronoi cells are translations and rotations of one fixed geometry. This simplifies the analysis, but does not enable it: The conclusions of this section apply to any crystalline material.

\begin{figure*}
\centering
\begin{tabular}{c@{}c@{}c@{}c@{}c@{}c@{}c@{}c@{}c@{}c@{}c}
0K & 300K & 600K & 900K & 1200K & 1500K & 1800K & 2100K & 2400K & 2700K & 3000K \\
\includegraphics[width=0.09\textwidth, trim={3cm 2.5cm 3cm 2.5cm}, clip]{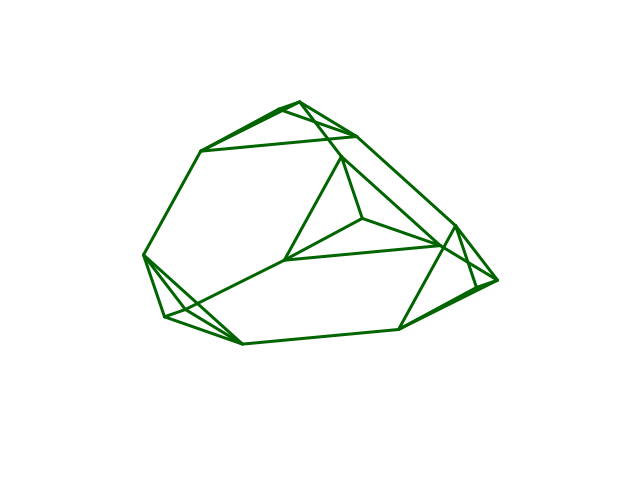} &
\includegraphics[width=0.09\textwidth, trim={3cm 2.5cm 3cm 2.5cm}, clip]{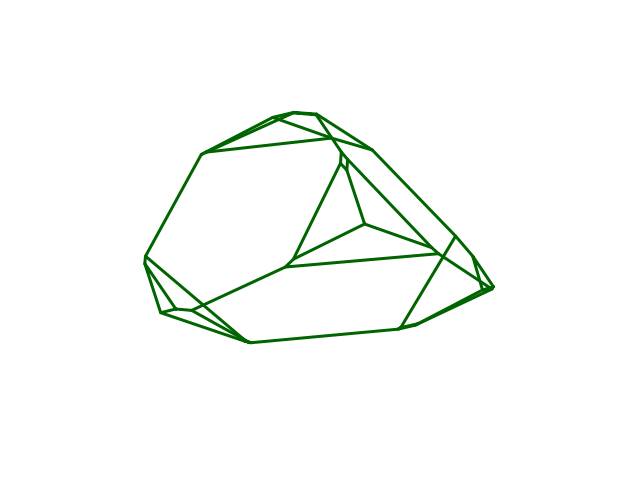} &
\includegraphics[width=0.09\textwidth, trim={3cm 2.5cm 3cm 2.5cm}, clip]{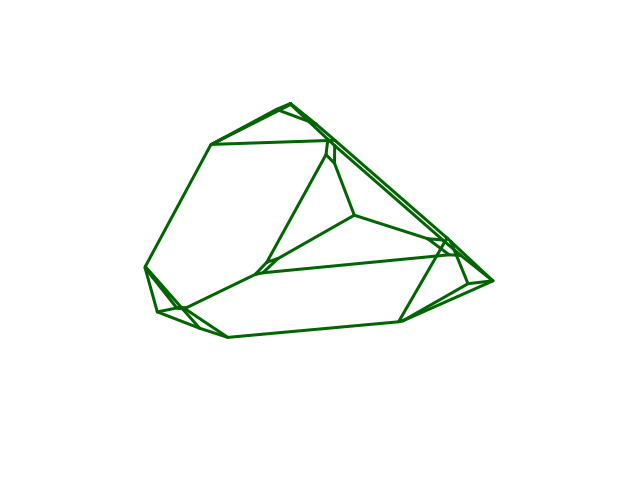} &
\includegraphics[width=0.09\textwidth, trim={3cm 2.5cm 3cm 2.5cm}, clip]{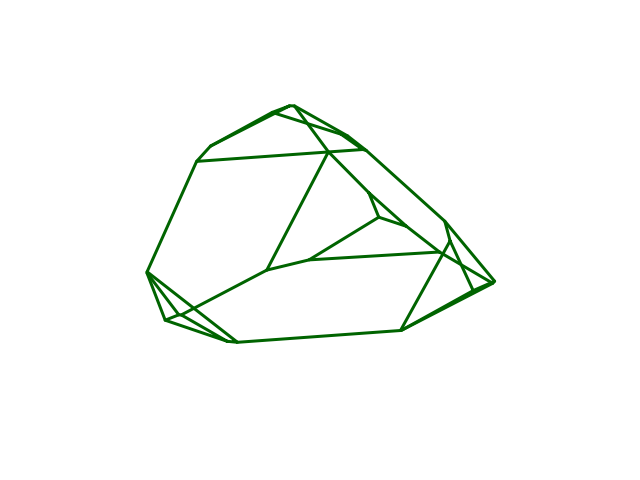} &
\includegraphics[width=0.09\textwidth, trim={3cm 2.5cm 3cm 2.5cm}, clip]{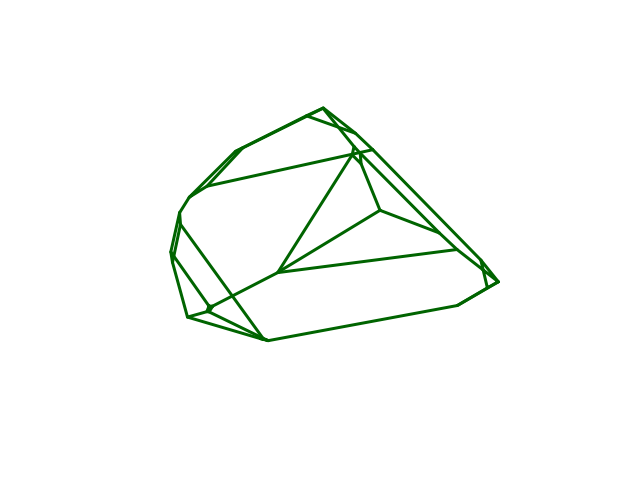} &
\includegraphics[width=0.09\textwidth, trim={3cm 2.5cm 3cm 2.5cm}, clip]{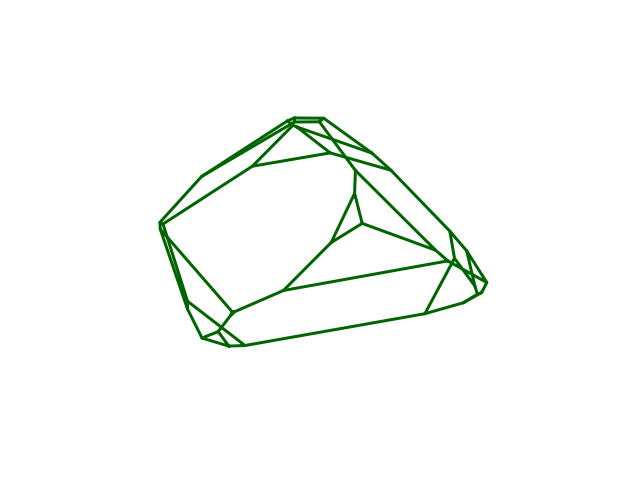} & 
\includegraphics[width=0.09\textwidth, trim={3cm 2.5cm 3cm 2.5cm}, clip]{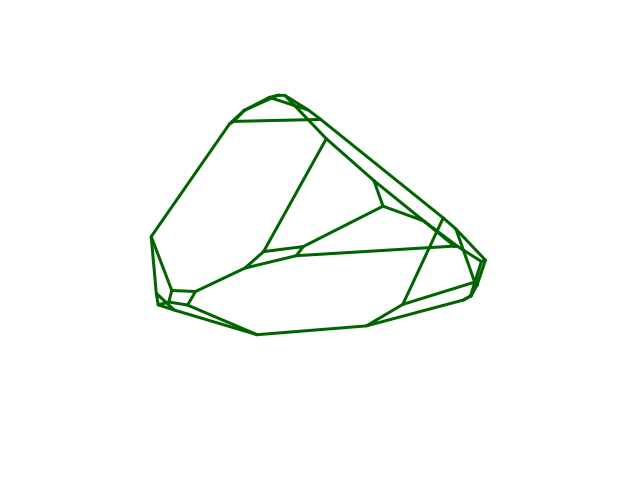} &
\includegraphics[width=0.09\textwidth, trim={3cm 2.5cm 3cm 2.5cm}, clip]{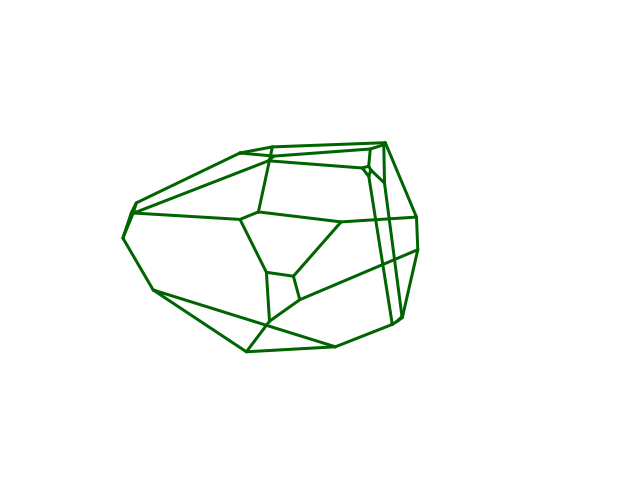} &
\includegraphics[width=0.09\textwidth, trim={3cm 2.5cm 3cm 2.5cm}, clip]{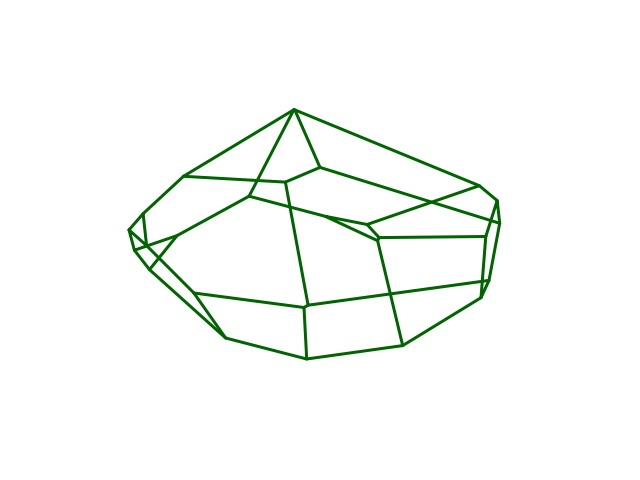} &
\includegraphics[width=0.09\textwidth, trim={3cm 2.5cm 3cm 2.5cm}, clip]{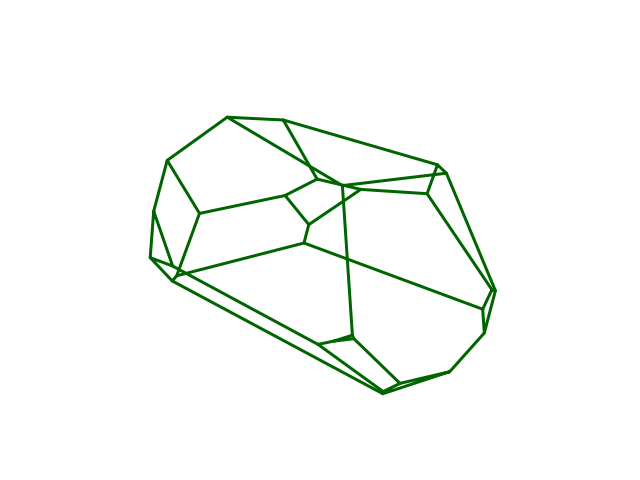} &
\includegraphics[width=0.09\textwidth, trim={3cm 2.5cm 3cm 2.5cm}, clip]{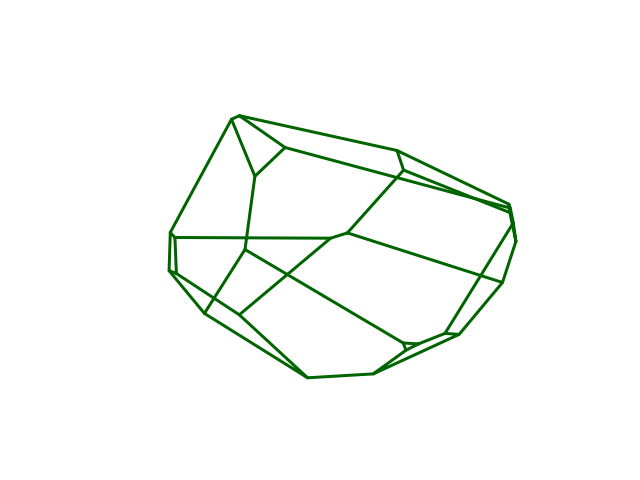} \\
\includegraphics[width=0.09\textwidth, trim={3cm 2.5cm 3cm 2.5cm}, clip]{Fig/0_g.png} &
\includegraphics[width=0.09\textwidth, trim={3cm 2.5cm 3cm 2.5cm}, clip]{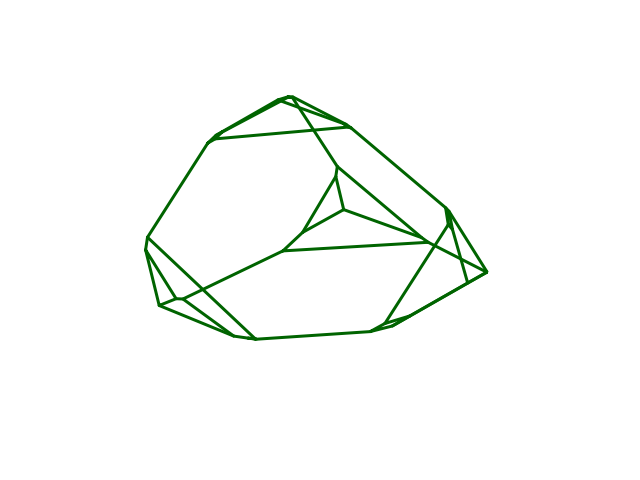} &
\includegraphics[width=0.09\textwidth, trim={3cm 2.5cm 3cm 2.5cm}, clip]{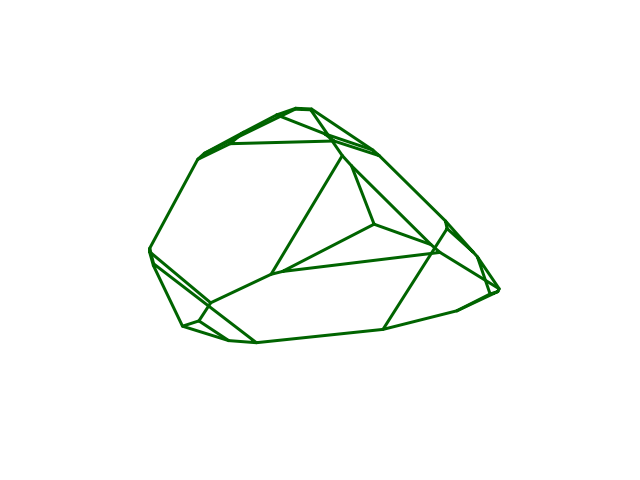} &
\includegraphics[width=0.09\textwidth, trim={3cm 2.5cm 3cm 2.5cm}, clip]{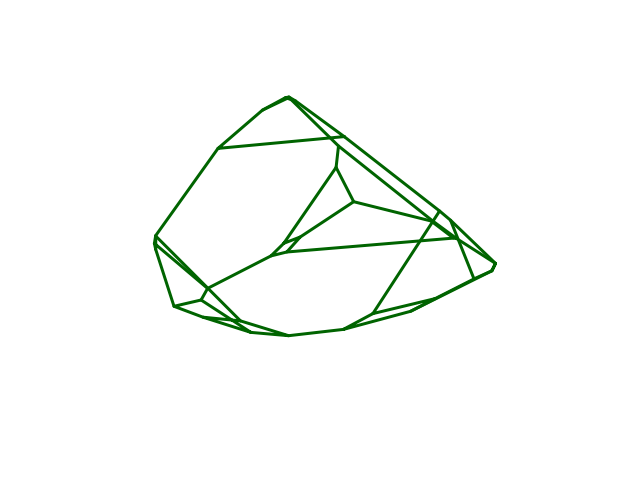} &
\includegraphics[width=0.09\textwidth, trim={3cm 2.5cm 3cm 2.5cm}, clip]{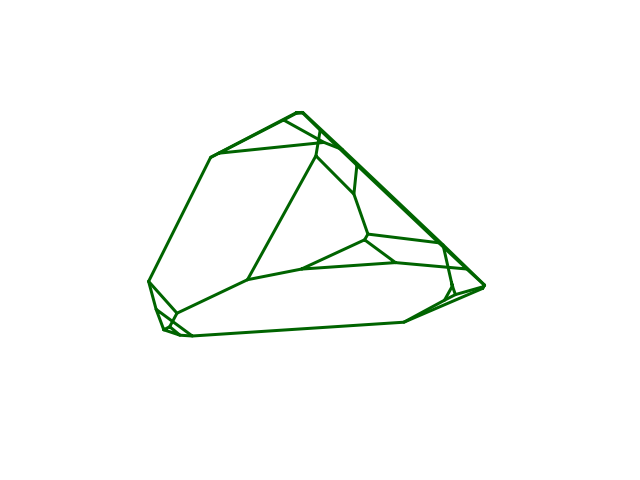} &
\includegraphics[width=0.09\textwidth, trim={3cm 2.5cm 3cm 2.5cm}, clip]{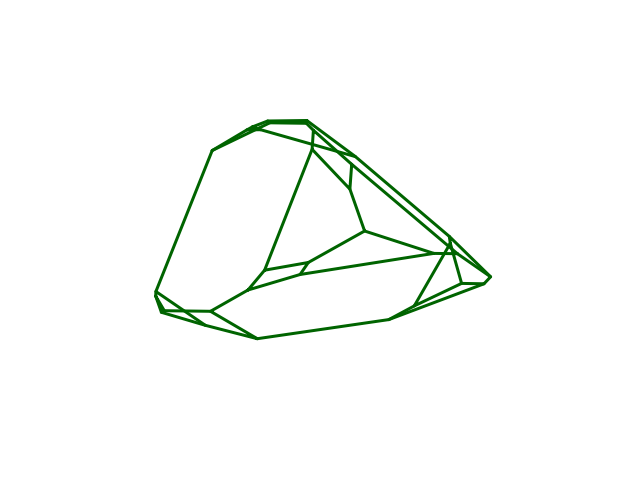} &
\includegraphics[width=0.09\textwidth, trim={3cm 2.5cm 3cm 2.5cm}, clip]{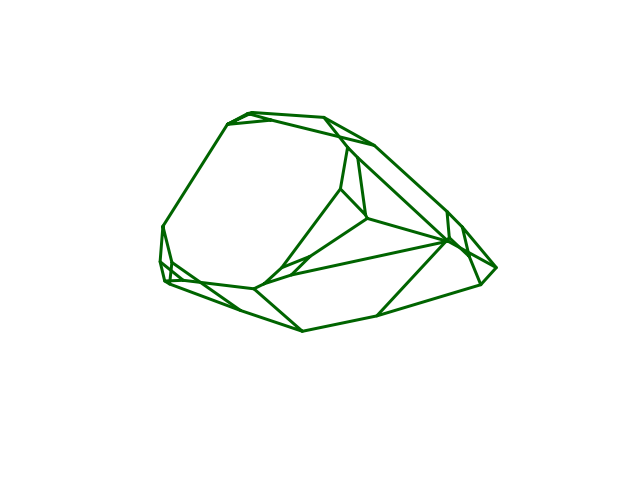} &
\includegraphics[width=0.09\textwidth, trim={3cm 2.5cm 3cm 2.5cm}, clip]{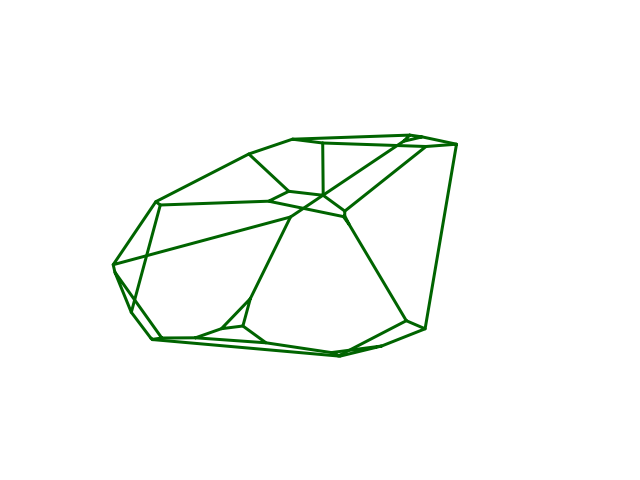} &
\includegraphics[width=0.09\textwidth, trim={3cm 2.5cm 3cm 2.5cm}, clip]{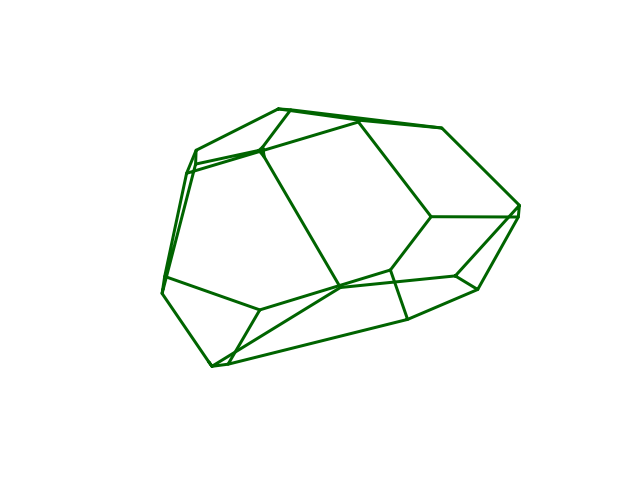} &
\includegraphics[width=0.09\textwidth, trim={2.4cm 2.5cm 3.6cm 2.5cm}, clip]{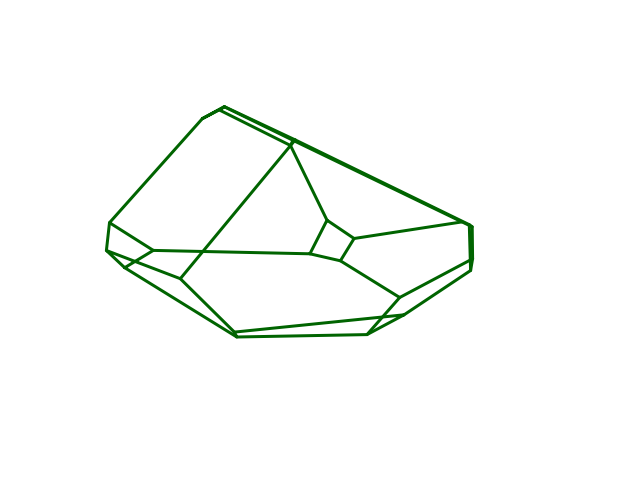} &
\includegraphics[width=0.09\textwidth, trim={3cm 2.5cm 3cm 2.5cm}, clip]{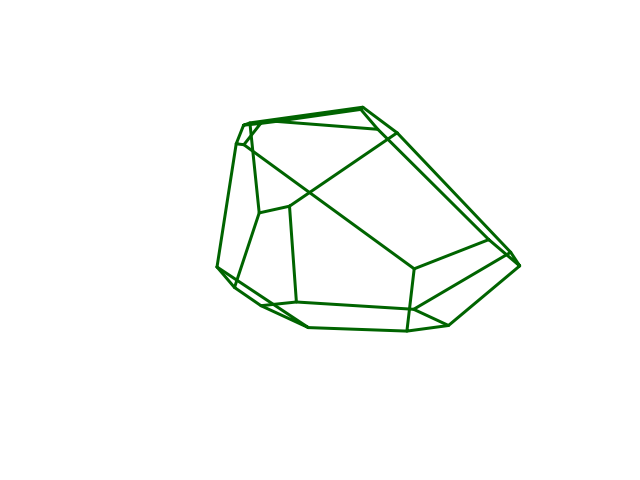} \\
\includegraphics[width=0.09\textwidth, trim={3cm 2.5cm 3cm 2.5cm}, clip]{Fig/0_g.png} &
\includegraphics[width=0.09\textwidth, trim={3cm 2.5cm 3cm 2.5cm}, clip]{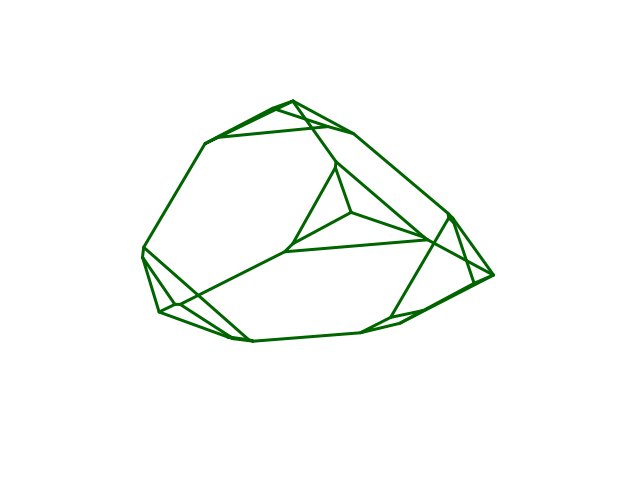} &
\includegraphics[width=0.09\textwidth, trim={3cm 2.5cm 3cm 2.5cm}, clip]{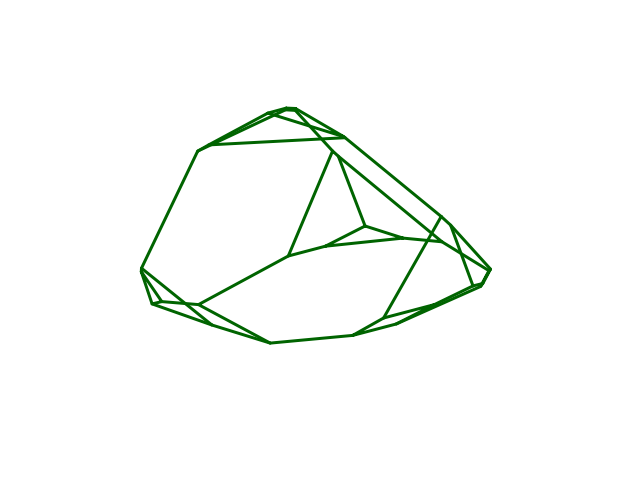} &
\includegraphics[width=0.09\textwidth, trim={3cm 2.5cm 3cm 2.5cm}, clip]{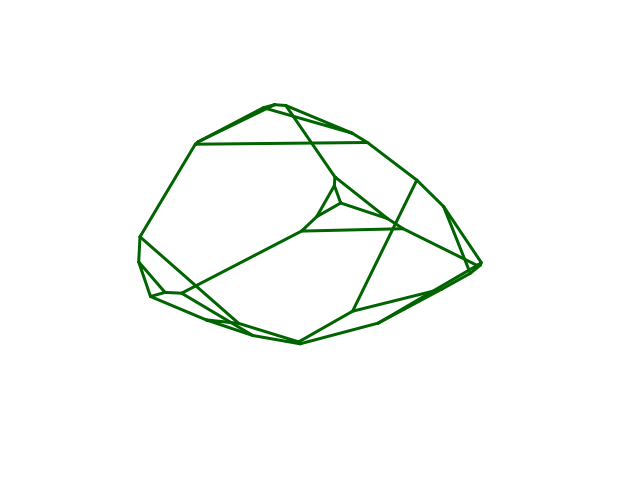} &
\includegraphics[width=0.09\textwidth, trim={3cm 2.5cm 3cm 2.5cm}, clip]{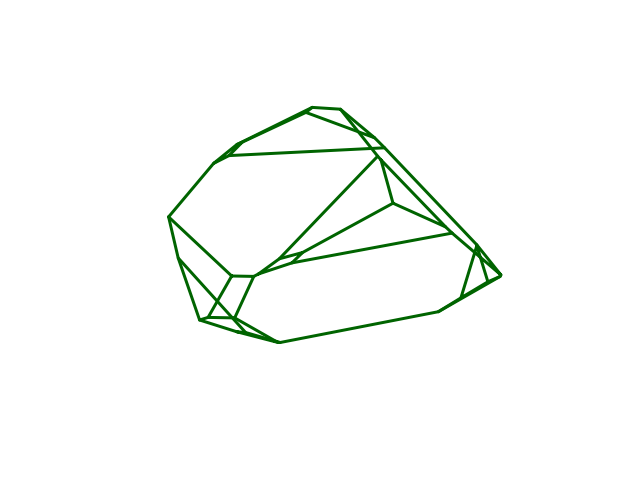} &
\includegraphics[width=0.09\textwidth, trim={3cm 2.5cm 3cm 2.5cm}, clip]{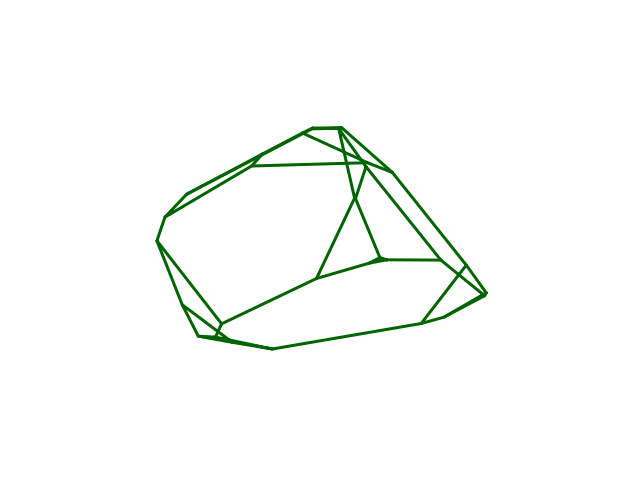} &
\includegraphics[width=0.09\textwidth, trim={3cm 2.5cm 3cm 2.5cm}, clip]{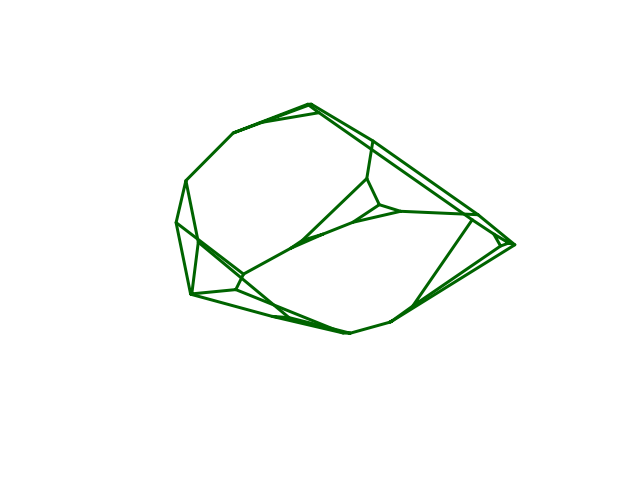} &
\includegraphics[width=0.09\textwidth, trim={3cm 2.5cm 3cm 2.5cm}, clip]{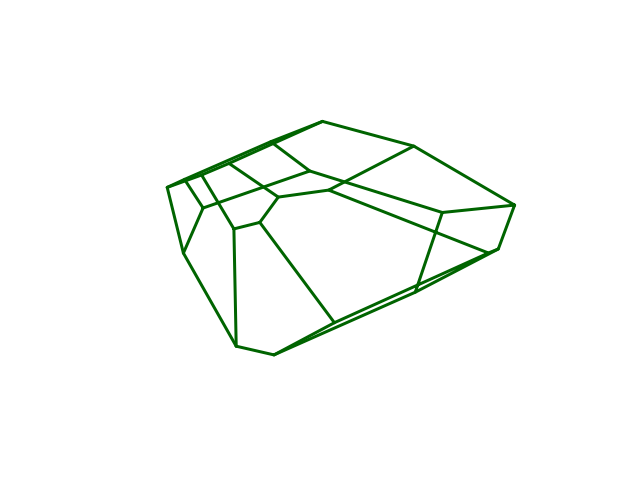} &
\includegraphics[width=0.09\textwidth, trim={3cm 2.5cm 3cm 2.5cm}, clip]{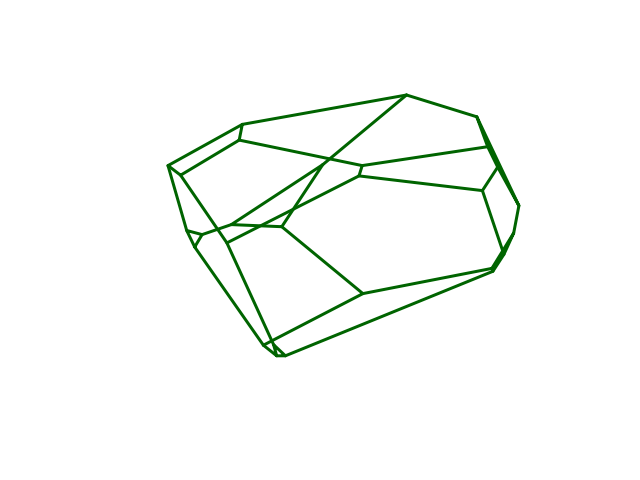} &
\includegraphics[width=0.09\textwidth, trim={3cm 2.5cm 3cm 2.5cm}, clip]{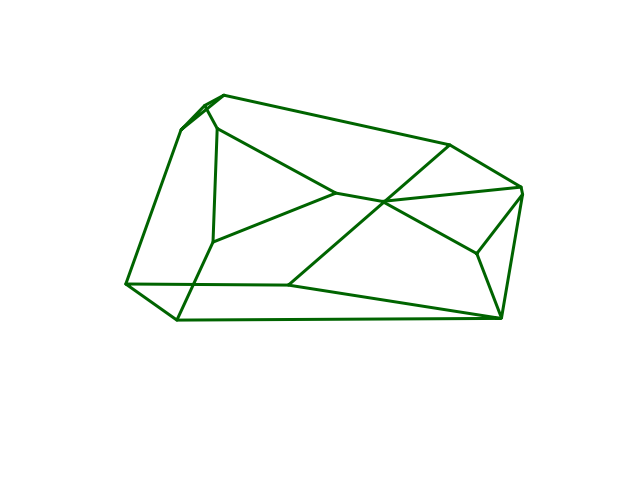} &
\includegraphics[width=0.09\textwidth, trim={3cm 2.5cm 3cm 2.5cm}, clip]{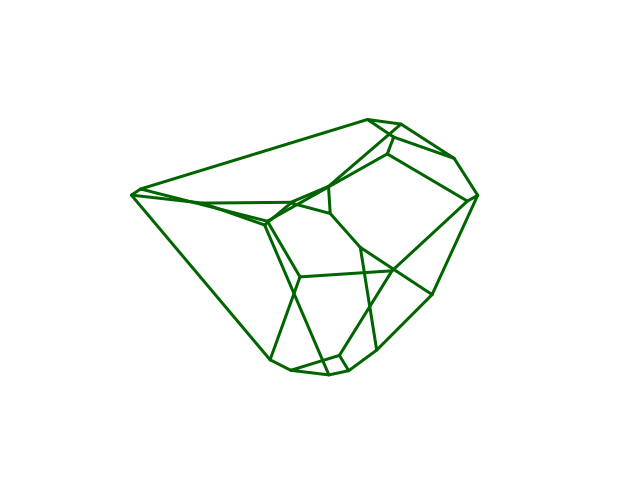}
\end{tabular}
\caption{The Voronoi cell of an silicon atom, as sampled from temporal orbits computed at specified temperatures and fixed volume using the second-generation Car-Parrinello FPMD method \cite{CPMD2007, kuhne2014second}.
}
\label{Fig:VoronoiDynamics}
\end{figure*}

Now, as one can see, the 0K-Voronoi cells in Fig.~\ref{Fig:SiOverview}(a) have 16 facets, which will require a prohibitive number of 48 parameters to quantify. Furthermore, under thermal fluctuations, the corners of the cells are unstable and additional facets appear. Besides the increased number of facets, the fluctuations in the number of facets complicate the task of quantifying the geometries of the Voronoi cells. To make these statements more quantitative, we sampled in Fig.~\ref{Fig:VoronoiDynamics} the geometry of one Voronoi cell during FPMD simulations performed at the seen temperatures, ranging from $300$K to $3000$K. A dramatic change in the overall geometry can be witnessed between 1800K - 2100K, which can be associated with the solid-liquid phase transition.\footnote{It is well known that the melting temperature of silicon is over-estimated by the type of FPMD simulations employed here \cite{sugino1995ab}.} One key observation is that, below the melting temperature, the large features of the Voronoi cells are stable, which immediately prompts the question of whether the smaller and wildly fluctuating features are actually needed when encoding the atomic configurations.

To answer it, we turn to the statistics presented in Fig.~\ref{Fig:VoronoiStat}. First, note that the histograms of the number of facets show an abrupt and dramatic change between the temperatures $1800$K to $2100$K, a clear indication of the first-order solid-liquid phase transition in silicon \cite{sugino1995ab}. Note that at temperatures below the melting line, there are situations that occur with finite probability, where the Voronoi cells display as many as 25 facets! The most important data, however, is contained in the second panel, which shows the histograms of the area of the facets at the given temperatures. At low temperatures, we see two very well-defined peaks, one at the right side coming from the four large facets and the other one coming from the fluctuations of the remaining twelve smaller facets. As already mentioned above, even smaller facets show up in the samples, with their histograms peaked at zero area. Now, the interesting fact in Fig.~\ref{Fig:VoronoiStat} is that the first peak disappears at higher temperatures, which are still well inside the crystalline phase, but the last peak endures all the way to the solid-liquid phase transition! By examining both families of histograms shown in Fig.~\ref{Fig:VoronoiStat}, we learned that this last peak comes with probability one from {\it exactly} four large facets of the Voronoi cells, which correspond to the planes separating pairs of first nearest-neighboring atoms. This empirical observation assures us that an atom of crystalline silicon continues to have precisely four first nearest-neighbors, all the way to the melting line.

Before drawing practical consequences of the above conclusions, let us clarify an important point that will highlight again the importance of Voronoi cells. Indeed, based on the above observations, it might seem that we can examine directly the nearest-neighboring pairs of atoms and avoid entirely the Voronoi cells. However, the notion of nearest-neighboring atoms, while useful, is very imprecise in the context of a strongly thermally fluctuating lattice. For example, this notion might lose its meaning entirely for crystals with more complex lattice structures, where first and second nearest-neighboring atoms cannot be separated due to thermal fluctuations. On the other hand, we can always examine the dynamics and statistics of the Voronoi cells, and determine the facets of the Voronoi cells that remain stable for temperatures up to either melting or structural phase transitions. By doing so, we can identify the pairs of atoms that, as we will conjecture, embodies the essence of a specific crystalline phase. For the silicon crystal, it so happens that these pairs correspond to the (clearly defined!) first nearest-neighboring pairs of atoms.

The above observation enables us to dramatically simplify the description of the atomic configurations of the silicon crystal. Indeed, suppose that, out of the complex geometry of the Voronoi cells, we only retain the data pertaining to the four largest facets, hence the four vectors $\bm v_n$ introduced above and featured in Fig.~\ref{Fig:SiOverview}(b). It is important at this point that the crystal is fully assembled and not broken down into pieces, as we previously did, in order to properly fix the {\it orientations} of the cells. Now, suppose this data was collected from each Voronoi cell and was placed in a file without any reference to the original locations of the Voronoi cells. Can we uniquely reproduce the crystal from this file? The answer is yes for the crystalline phase! For this, we can start from any line of such a file, from where we can read four vectors encoding the information we just discussed. We place a point anywhere in space, say at the origin of $\mathbb R^3$ and draw the four vectors, which we will call $\bm v_n(1)$. At the same time, we erase the line corresponding to atom 1 from the data file. Then we cycle through the remaining data until we find a line that contains a vector $\bm v$ that is exactly opposite to one of the vectors $\bm v_n(1)$.\footnote{Given the thermal displacements of atoms, there will be zero probability to find another vector in our table that matches $\bm v_n(1)$ with machine precision.} At that instance, we have found a first nearest-neighbor of atom 1, which we place at its rightful location at $2\bm v_n(1)$ together with its four corresponding vectors, which we call $\bm v_n(2)$. At the same time, we erase the line corresponding to atom 2 from the data file. Cycling again through the remaining data, we will find a line that contains one vector $\bm v'$ that is exactly opposite to one of the vectors $\bm v_n(1)$ or $\bm v_n(2)$, respectively. This is a first nearest-neighbor atom of the dimer made out of atoms 1 and 2. By iterating the procedure $k$ times, we grow a cluster of $k+1$ atoms by adding one first nearest-neighboring atom per iteration, and we verified that the procedure exhausts all atoms of a finite crystal after a finite number of steps. We arrived at our second main finding:

\begin{figure}
\centering
\includegraphics[width=0.24\textwidth,trim={0cm 0 0.cm 0},clip]{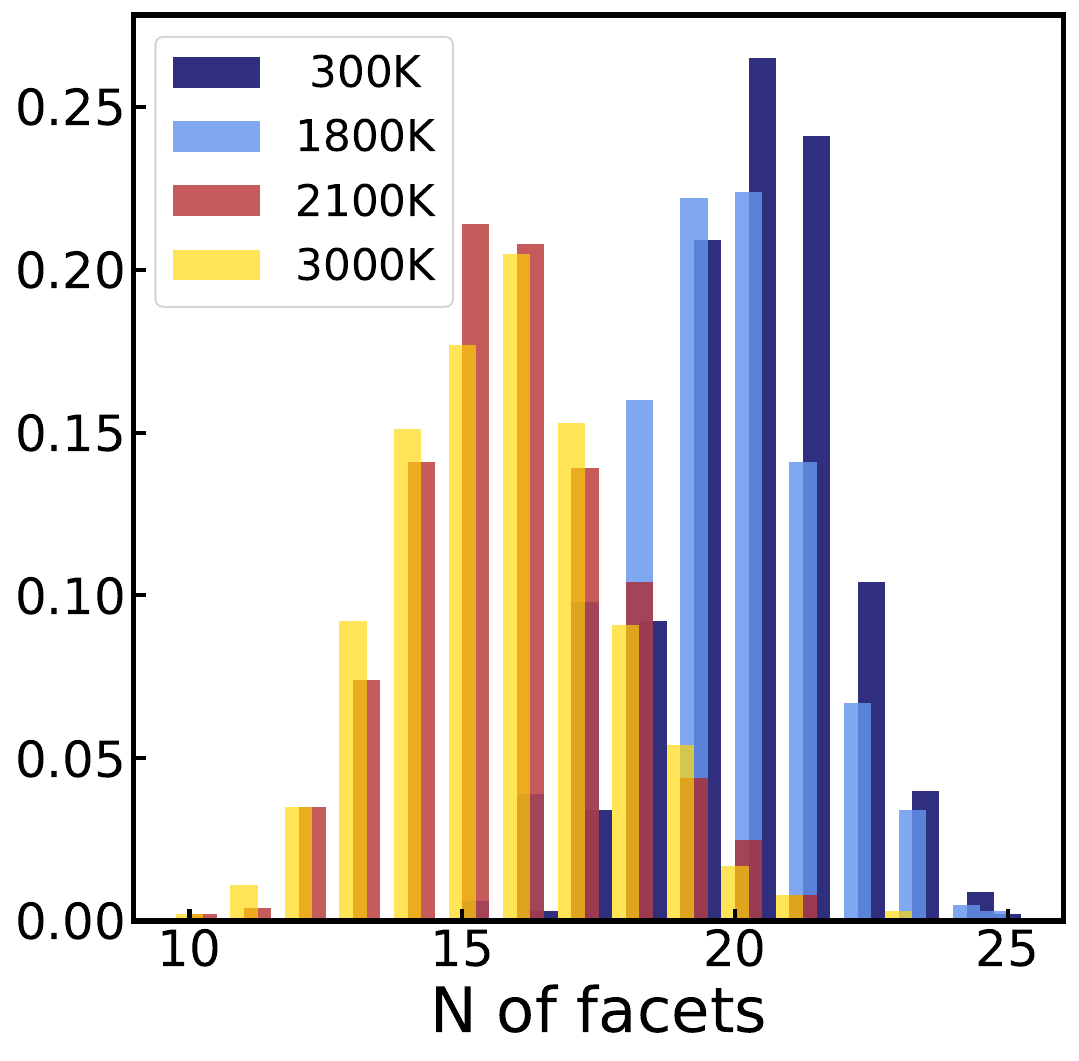}
\includegraphics[width=0.24\textwidth,trim={0cm 0 0.cm 0},clip]{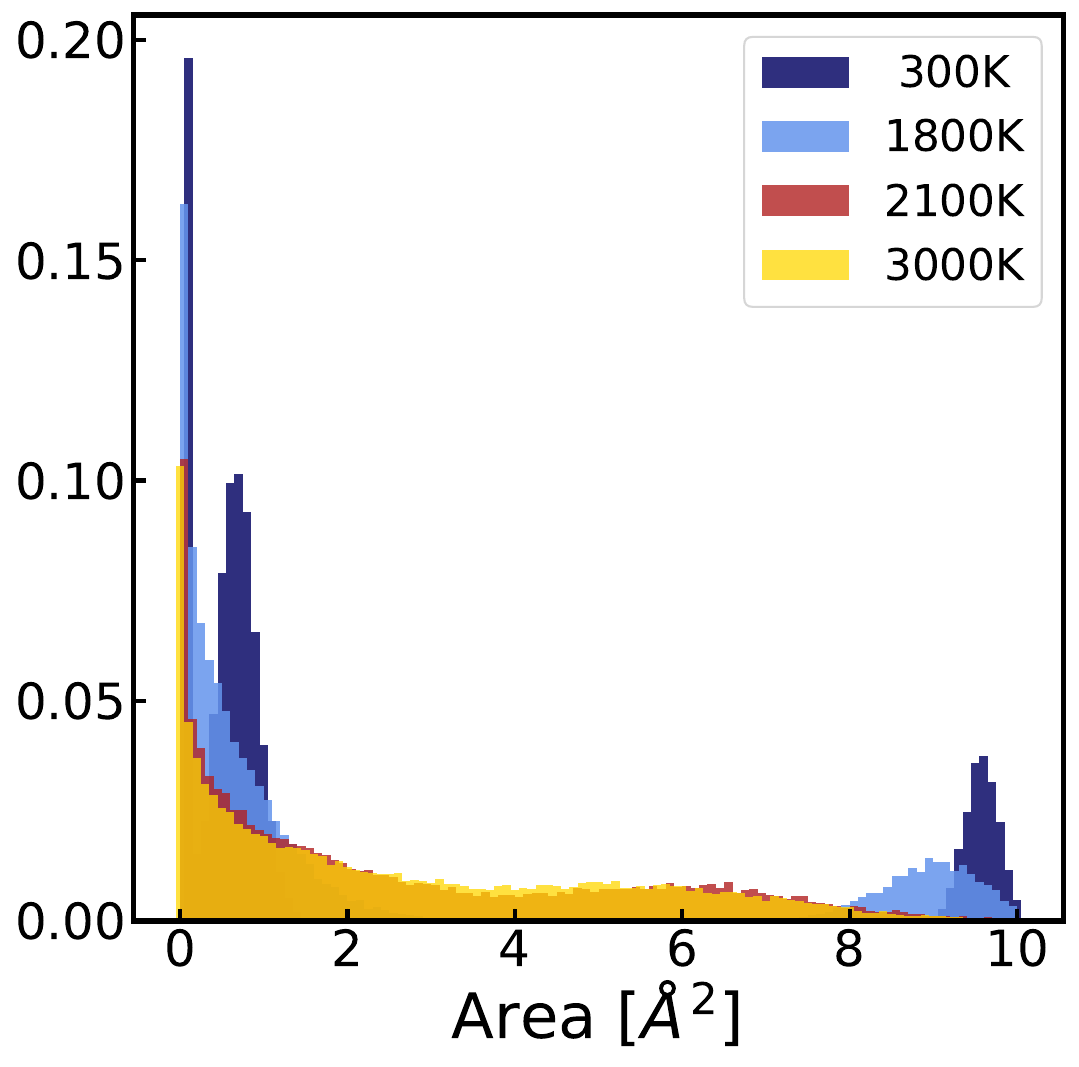}
\caption{Upon phase transition, the structure of a Voronoi cell experience a drastic change, as we demonstrate on the number of facets (\textit{left}) and total area (\textit{right}) of a cell. Below phase transition the Voronoi cell for a silicon atom is composed of 4 large and several small facets, whilst after the transition point we observe no dominant facets.
}
\label{Fig:VoronoiStat}
\end{figure}

{\bf Statement 2.} {\it For crystalline silicon, the Gibbs measure of the classical degrees of freedom is fully encoded in the data collected from the four largest Voronoi facets, which are stable to temperatures up to the melting line of the crystal.}  

We point out that our technique fails for the liquid phase, because the number and orientations of large facets are both strongly fluctuating, but there are other techniques based on Voronoi tessellations that cover the liquid phases (see \cite{bellissard2017anankeontheoryviscosityliquids}). However, our technique can be applied to any crystalline material, regardless of the complexity of its unit cell. Indeed, the dynamics and statistics of the Voronoi cells can be used to identify temperature-induced phase transitions, and to determine the facets of the Voronoi cells that remain stable up to that transition. One then only needs to check that the data associated with those facets is enough to reconstruct the crystal, by following the protocol we explained above. This touches on an issue of great importance in crystallography, namely, how to find a unique real-space marker of a crystalline phase of a compound with a complicated phase diagram, {\it e.g.} with multiple structural phase transitions? We believe we found a rigorous device to answer this question:

{\bf Conjecture.} {\it A crystalline phase of a condensed matter system is fully and uniquely encoded by the facets of the Voronoi cells that are stable throughout that specific part of the phase diagram.}

\begin{figure*}
\begin{adjustbox}{width=0.82\textwidth,center}
\begin{subfigure}[t]{.22\textwidth}
\centering
\includegraphics[width=\linewidth,trim={2.1cm 1.1cm 0cm 0},clip]{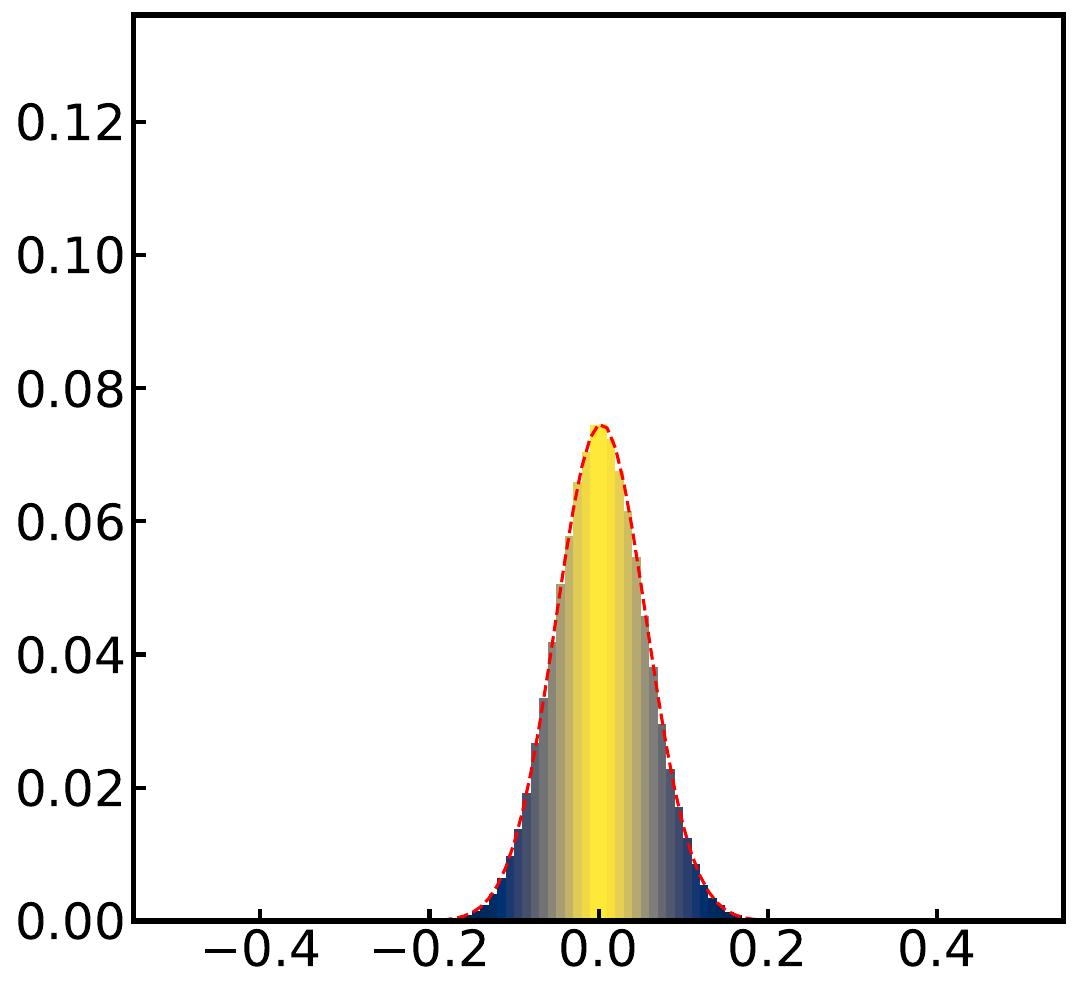}
\put(-37., 90){300 K}
\put(-62., 80){$\mu$ =  0.00298}
\put(-62., 70){$\sigma$ =  0.053}
\put(-128, -1.5){0.00}
\put(-128, 13.5){0.02}
\put(-128, 29){0.04}
\put(-128, 44.5){0.06}
\put(-128, 60){0.08}
\put(-128, 75.5){0.10}
\put(-128, 91){0.12}
\end{subfigure}\hfill
\begin{subfigure}[t]{.22\textwidth}
\centering
\includegraphics[width=\linewidth,trim={2.1cm 1.1cm 0cm 0},clip]{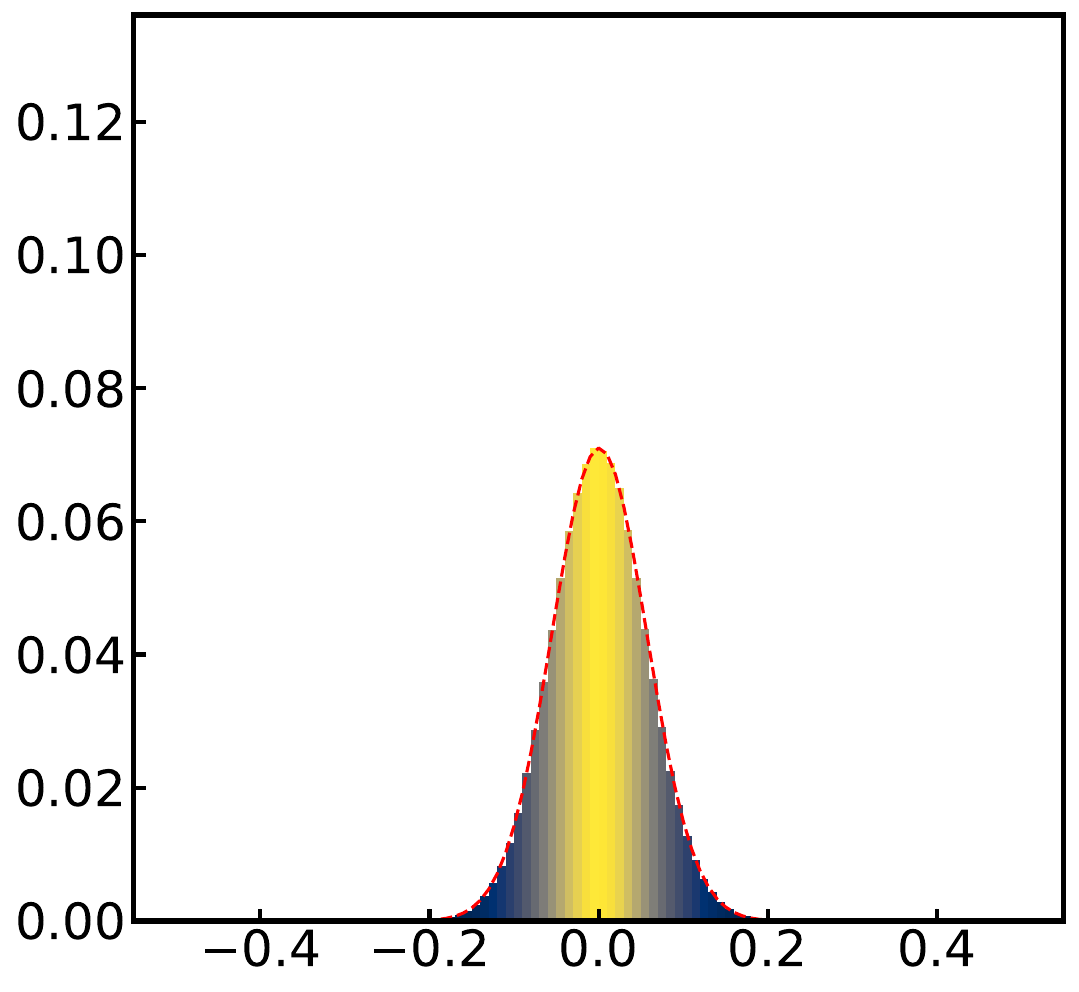}
\put(-37., 90){300 K}
\put(-62., 80){$\mu$ =  0.00086}
\put(-62., 70){$\sigma$ = 0.056}
\end{subfigure}\hfill
\begin{subfigure}[t]{.22\textwidth}
\centering
\includegraphics[width=\linewidth,trim={2.1cm 1.1cm 0cm 0},clip]{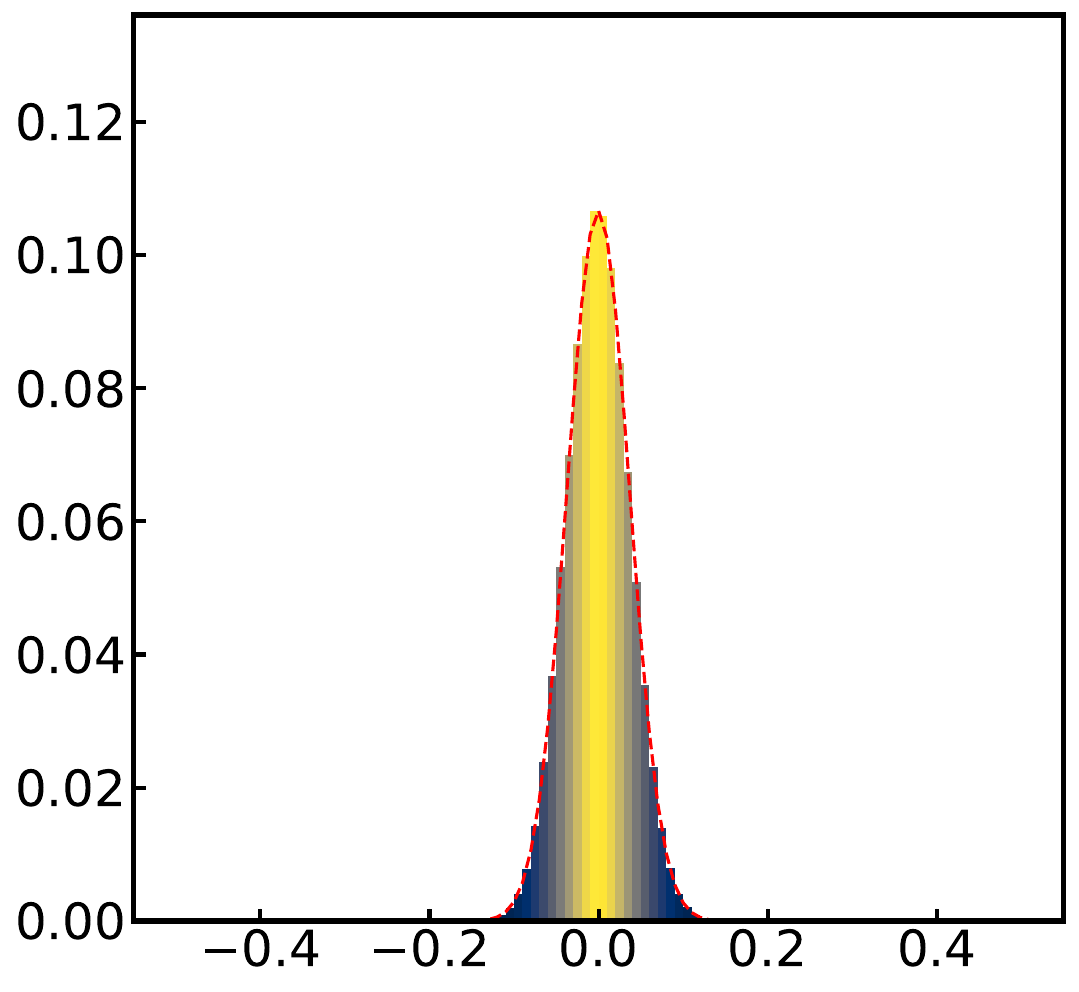}
\put(-37., 90){300 K}
\put(-62., 80){$\mu$ =  -0.00037}
\put(-62., 70){$\sigma$ =  0.037}
\end{subfigure}\hfill
\begin{subfigure}[t]{.22\textwidth}
\centering
\includegraphics[width=\linewidth,trim={2.1cm 1.1cm 0cm 0},clip]{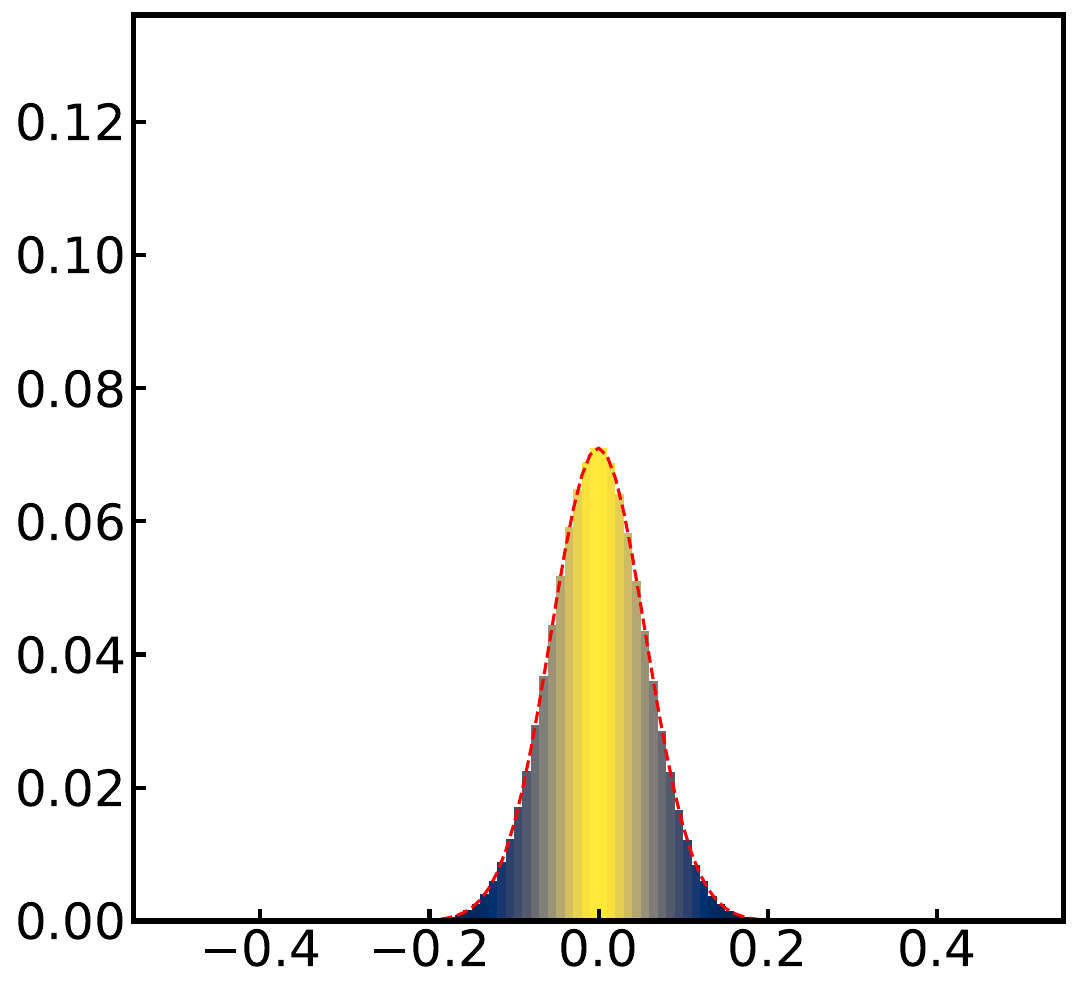}
\put(-37., 90){300 K}
\put(-62., 80){$\mu$ = -0.00042}
\put(-62., 70){$\sigma$ = 0.056}
\end{subfigure}\hfill

\end{adjustbox}
\begin{adjustbox}{width=0.82\textwidth,center}
\begin{subfigure}[t]{.22\textwidth}
\centering
\includegraphics[width=\linewidth,trim={2.1cm 1.1cm 0cm 0},clip]{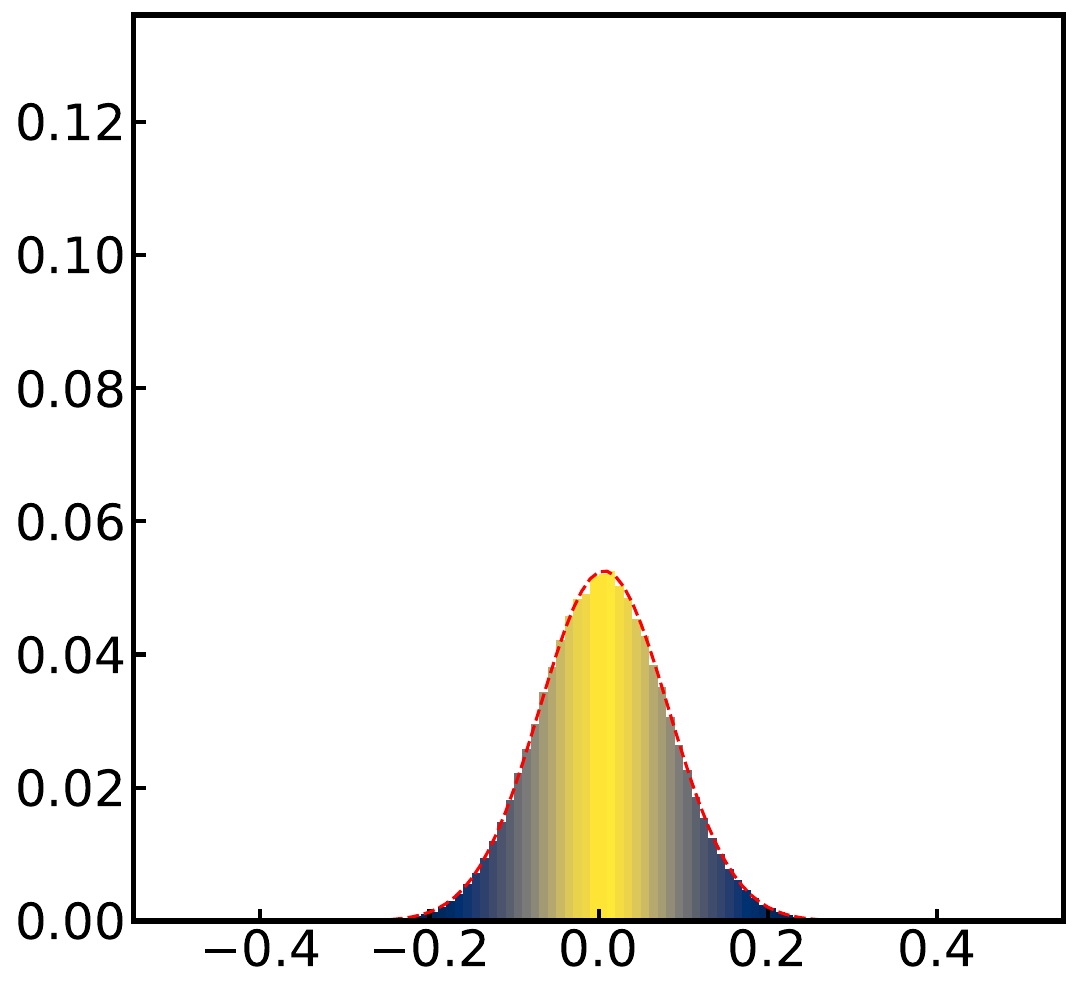}
\put(-37., 90){600 K}
\put(-62., 80){$\mu$ = 0.0065}
\put(-62., 70){$\sigma$ = 0.076}
\put(-128, -1.5){0.00}
\put(-128, 13.5){0.02}
\put(-128, 29){0.04}
\put(-128, 44.5){0.06}
\put(-128, 60){0.08}
\put(-128, 75.5){0.10}
\put(-128, 91){0.12}
\end{subfigure}\hfill
\begin{subfigure}[t]{.22\textwidth}
\centering
\includegraphics[width=\linewidth,trim={2.1cm 1.1cm 0cm 0},clip]{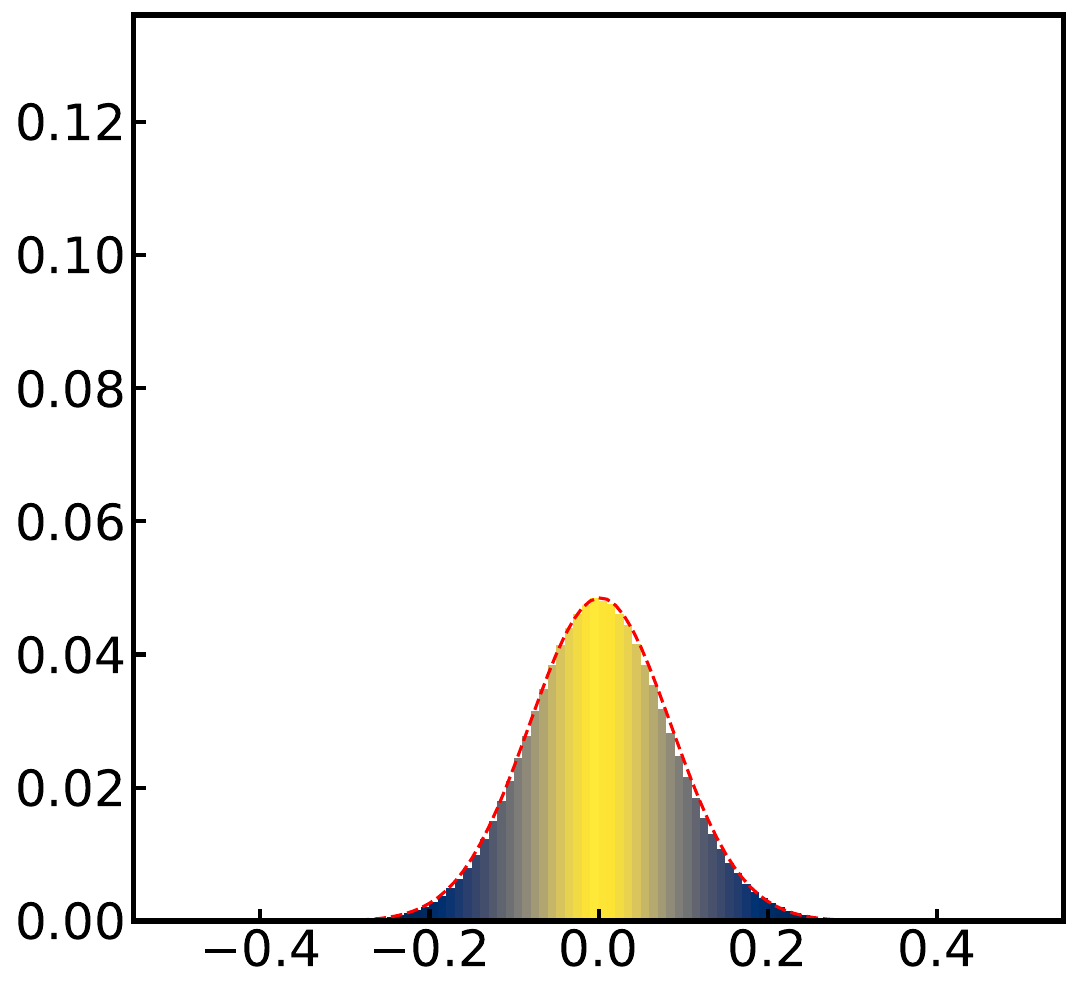}
\put(-37., 90){600 K}
\put(-62., 80){$\mu$ = 0.0018}
\put(-62., 70){$\sigma$ = 0.084}
\end{subfigure}\hfill
\begin{subfigure}[t]{.22\textwidth}
\centering
\includegraphics[width=\linewidth,trim={2.1cm 1.1cm 0cm 0},clip]{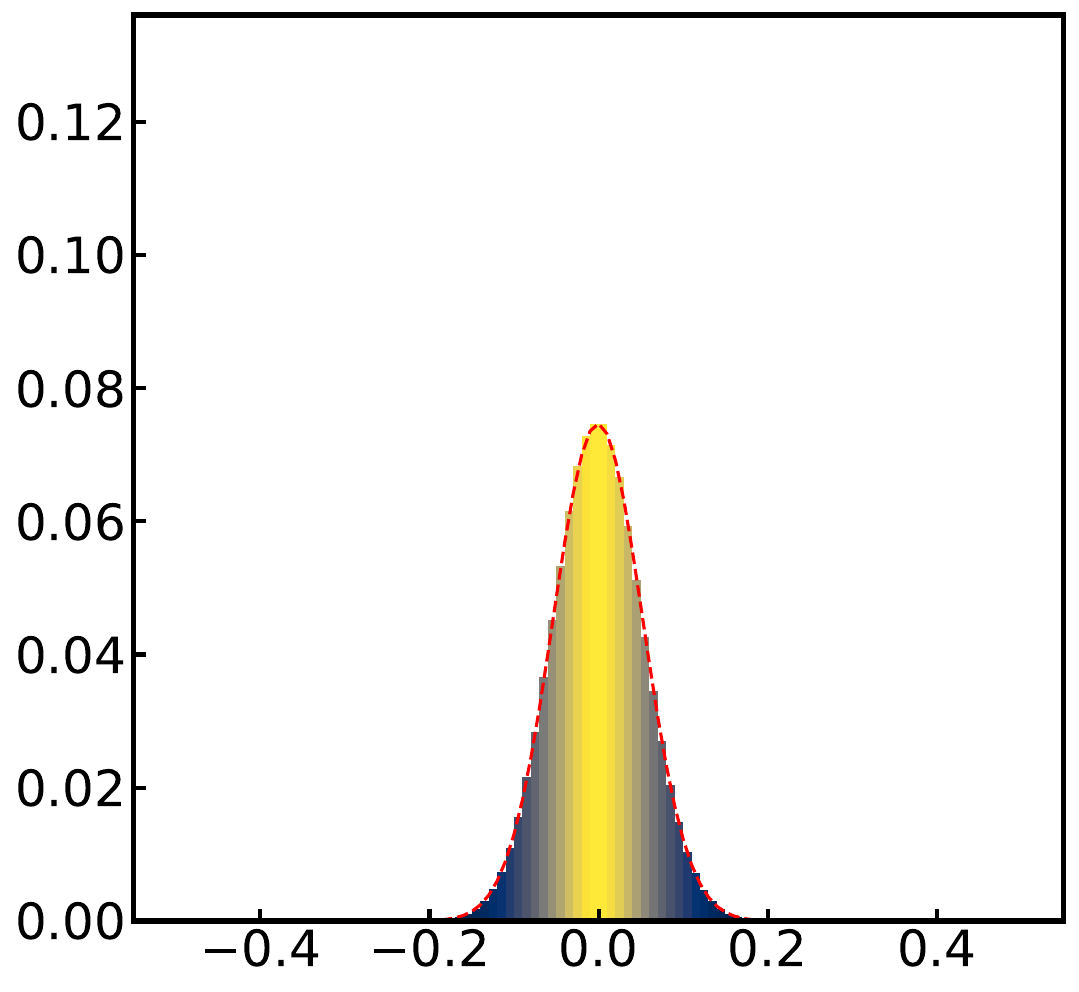}
\put(-37., 90){600 K}
\put(-62., 80){$\mu$ = -0.00082}
\put(-62., 70){$\sigma$ =  0.053}
\end{subfigure}\hfill
\begin{subfigure}[t]{.22\textwidth}
\centering
\includegraphics[width=\linewidth,trim={2.1cm 1.1cm 0cm 0},clip]{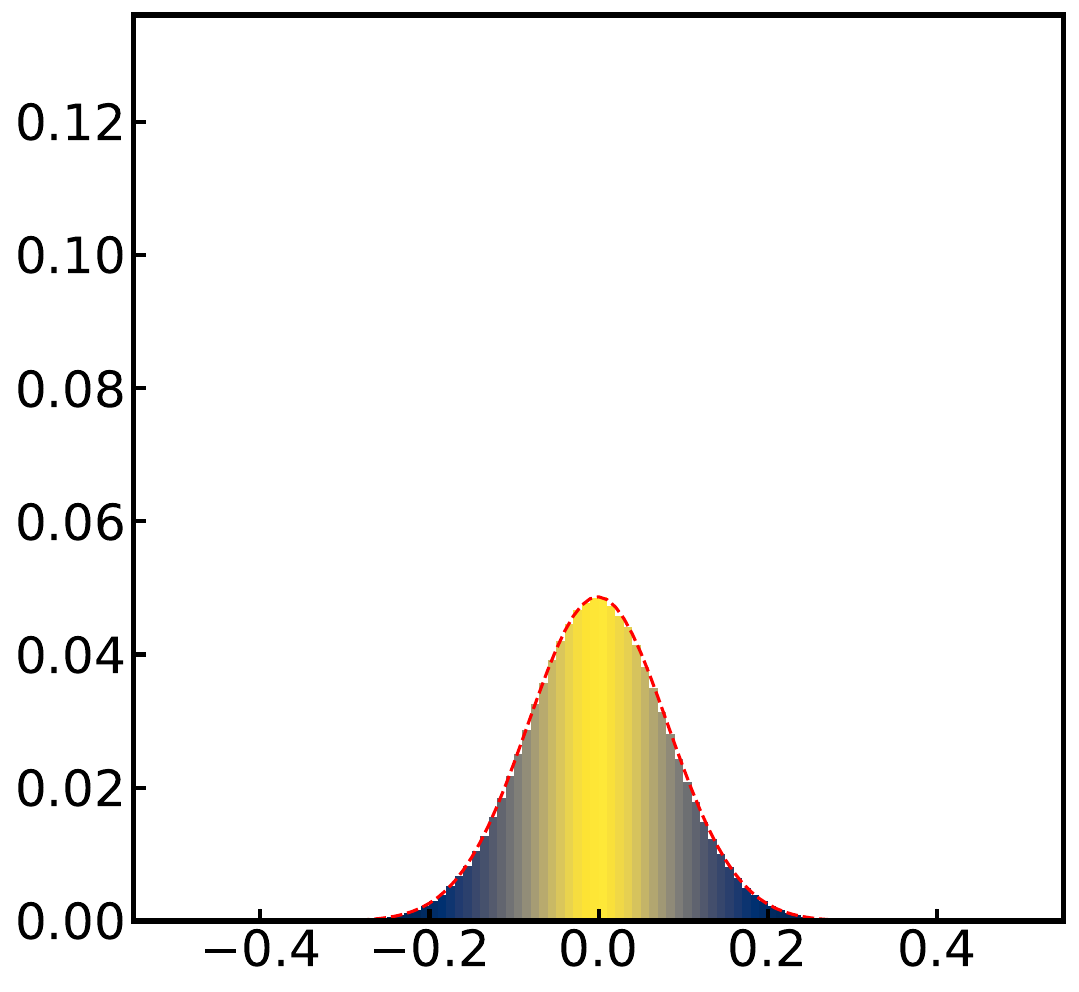}
\put(-37., 90){600 K}
\put(-62., 80){$\mu$ = -0.00093}
\put(-62., 70){$\sigma$ = 0.083}
\end{subfigure}\hfill
\end{adjustbox}
\begin{adjustbox}{width=0.82\textwidth,center}
\begin{subfigure}[t]{.22\textwidth}
\centering
\includegraphics[width=\linewidth,trim={2.1cm 1.1cm 0cm 0},clip]{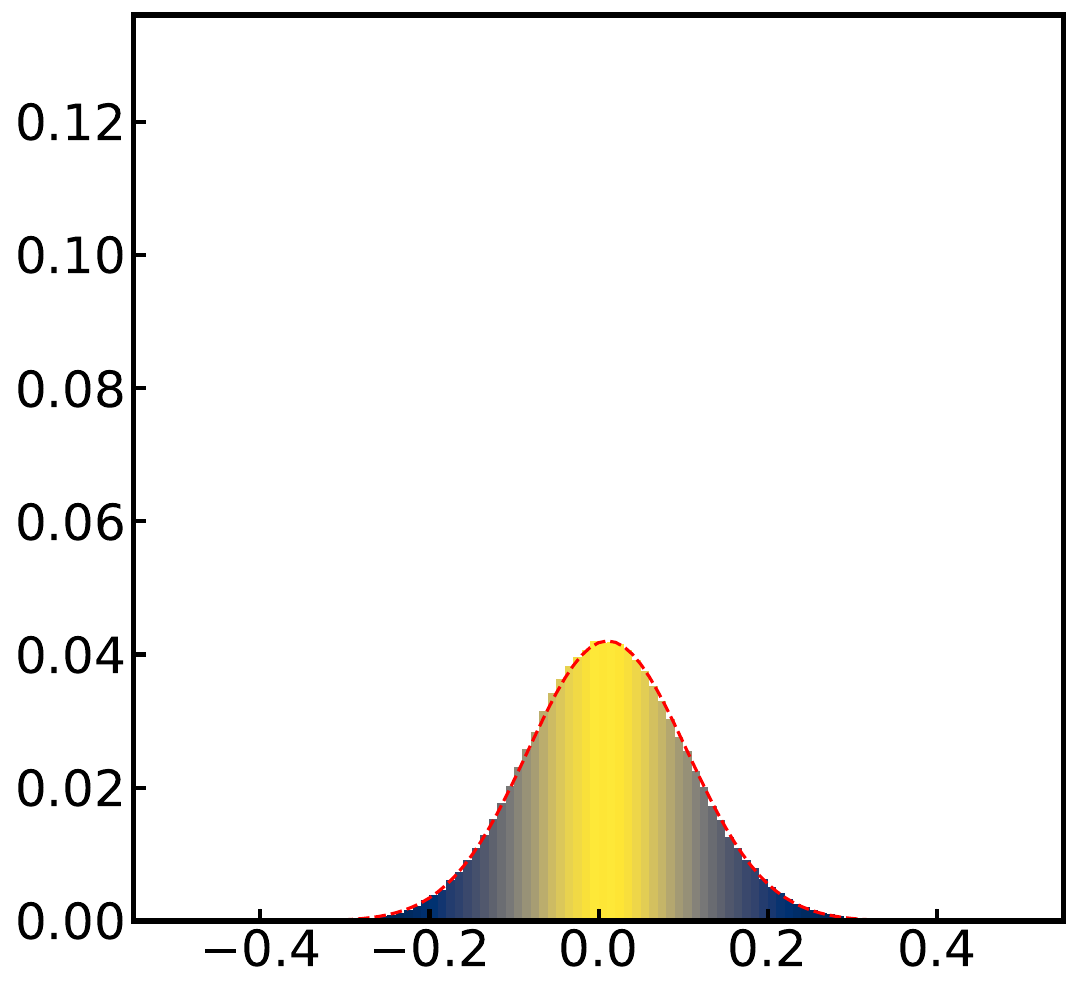}
\put(-37., 90){900 K}
\put(-62., 80){$\mu$ = 0.011}
\put(-62., 70){$\sigma$ = 0.094}
\put(-128, -1.5){0.00}
\put(-128, 13.5){0.02}
\put(-128, 29){0.04}
\put(-128, 44.5){0.06}
\put(-128, 60){0.08}
\put(-128, 75.5){0.10}
\put(-128, 91){0.12}
\end{subfigure}\hfill
\begin{subfigure}[t]{.22\textwidth}
\centering
\includegraphics[width=\linewidth,trim={2.1cm 1.1cm 0cm 0},clip]{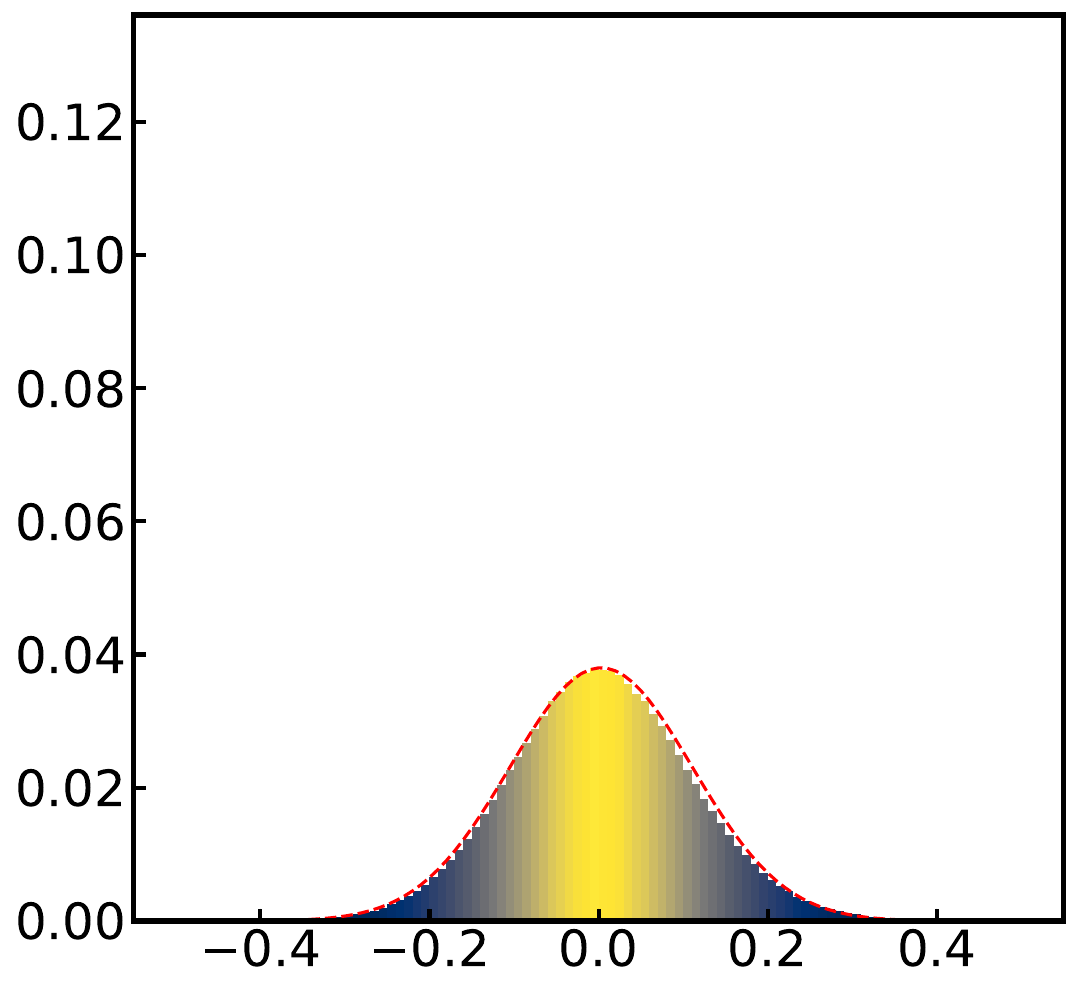}
\put(-37., 90){900 K}
\put(-55., 80){$\mu$ = 0.003}
\put(-55., 70){$\sigma$ = 0.11}
\end{subfigure}\hfill
\begin{subfigure}[t]{.22\textwidth}
\centering
\includegraphics[width=\linewidth,trim={2.1cm 1.1cm 0cm 0},clip]{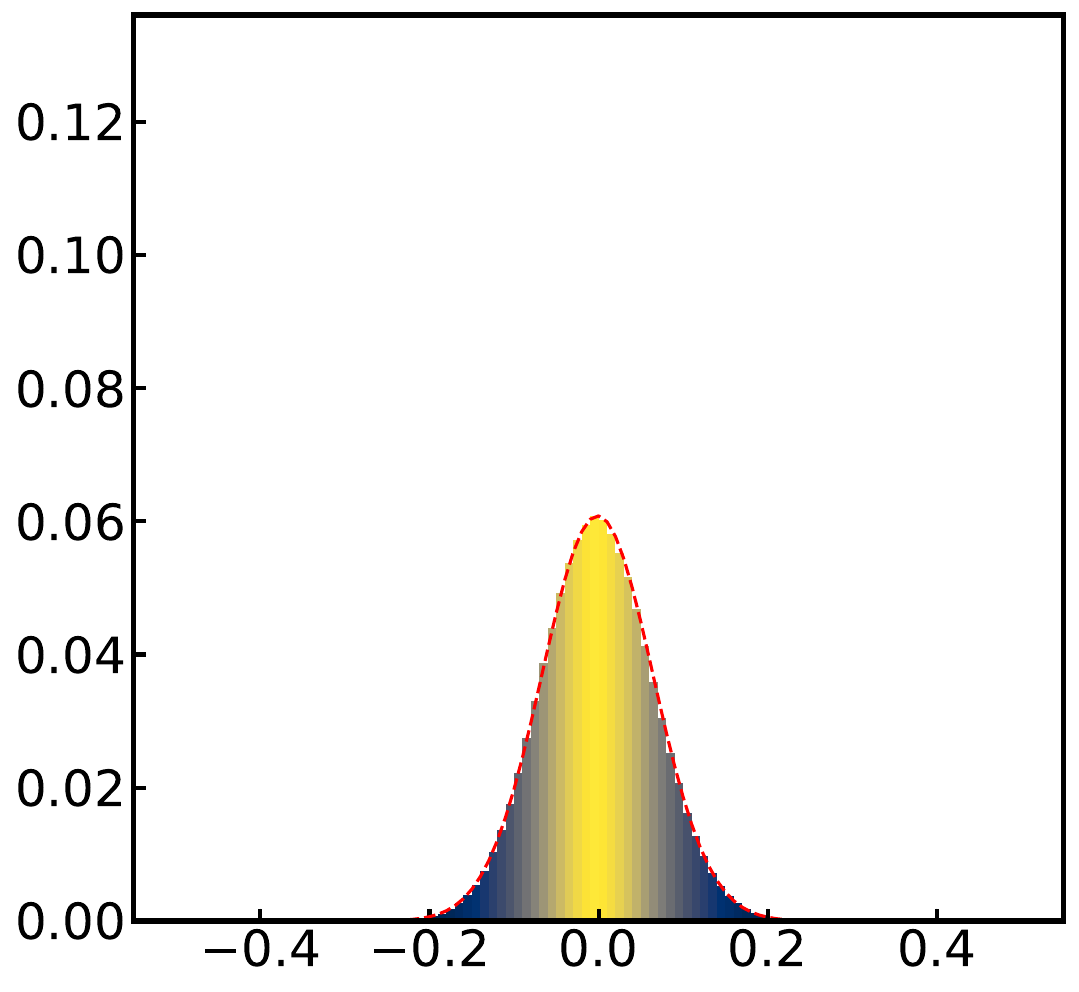}
\put(-37., 90){900 K}
\put(-62., 80){$\mu$ = -0.0014}
\put(-62., 70){$\sigma$ = 0.066}
\end{subfigure}\hfill
\begin{subfigure}[t]{.22\textwidth}
\centering
\includegraphics[width=\linewidth,trim={2.1cm 1.1cm 0cm 0},clip]{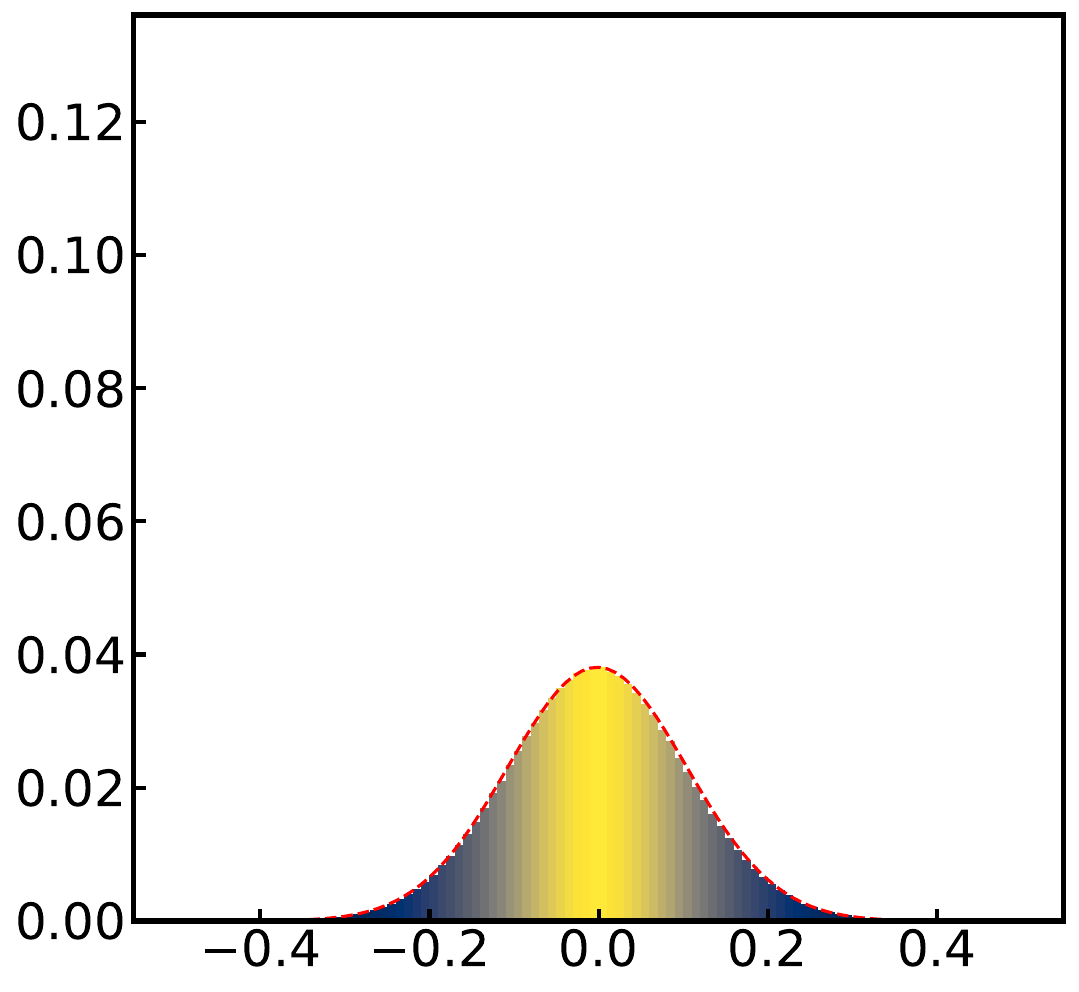}
\put(-37., 90){900 K}
\put(-62., 80){$\mu$ = -0.0015}
\put(-62., 70){$\sigma$ = 0.11}
\end{subfigure}\hfill
\end{adjustbox}
\begin{adjustbox}{width=0.82\textwidth,center}
\begin{subfigure}[t]{.22\textwidth}
\centering
\includegraphics[width=\linewidth,trim={2.1cm 1.1cm 0cm 0},clip]{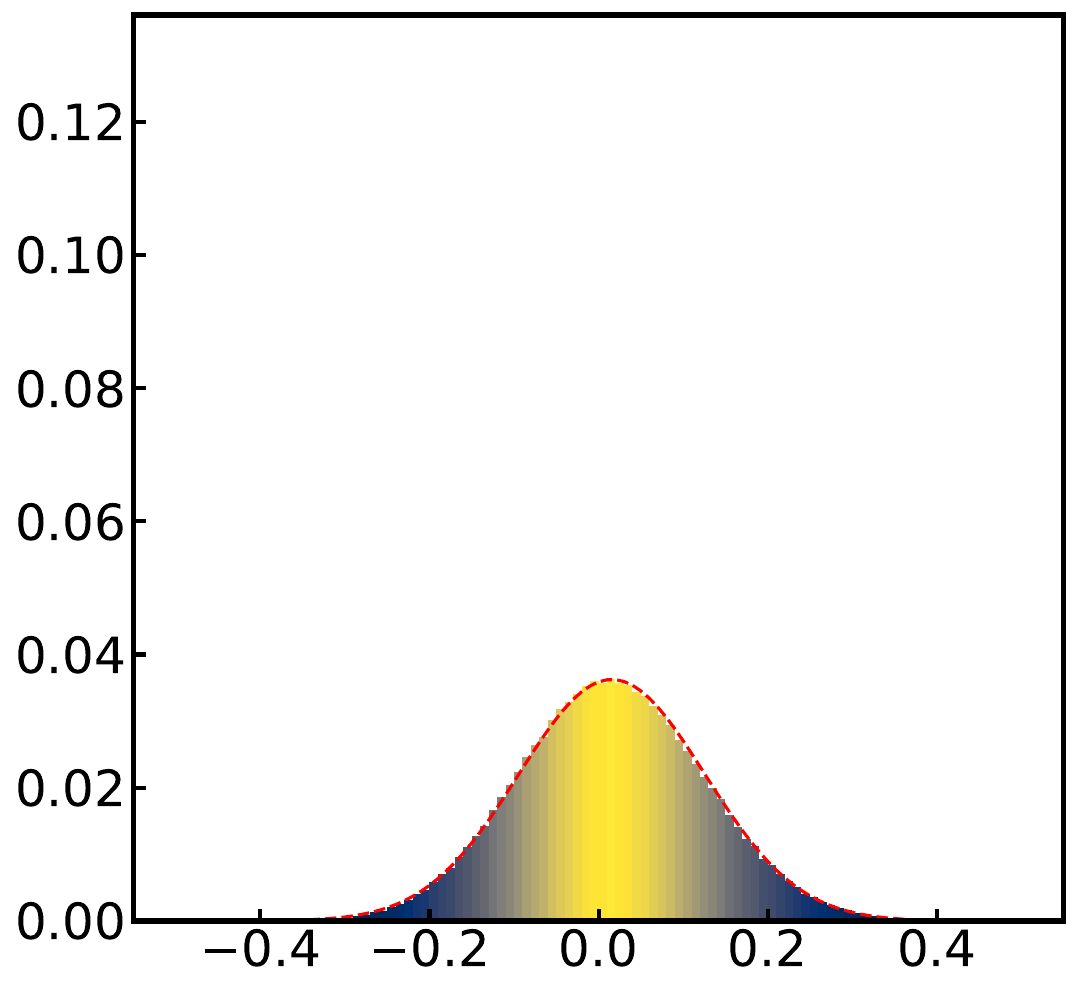}
\put(-37., 90){1200 K}
\put(-62., 80){$\mu$ = 0.016}
\put(-62., 70){$\sigma$ = 0.11}
\put(-128, -1.5){0.00}
\put(-128, 13.5){0.02}
\put(-128, 29){0.04}
\put(-128, 44.5){0.06}
\put(-128, 60){0.08}
\put(-128, 75.5){0.10}
\put(-128, 91){0.12}
\end{subfigure}\hfill
\begin{subfigure}[t]{.22\textwidth}
\centering
\includegraphics[width=\linewidth,trim={2.1cm 1.1cm 0cm 0},clip]{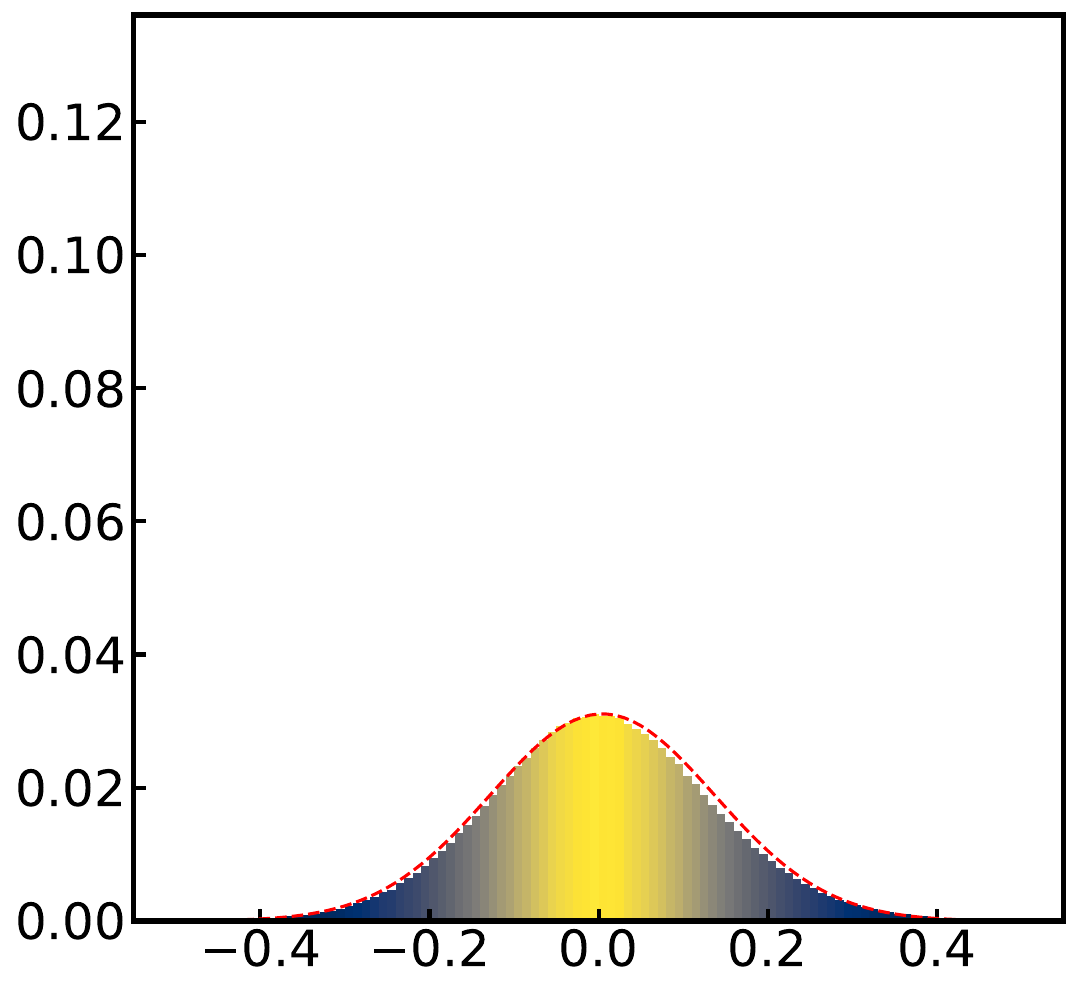}
\put(-37., 90){1200 K}
\put(-62., 80){$\mu$ = 0.0044}
\put(-62., 70){$\sigma$ = 0.13}
\end{subfigure}\hfill
\begin{subfigure}[t]{.22\textwidth}
\centering
\includegraphics[width=\linewidth,trim={2.1cm 1.1cm 0cm 0},clip]{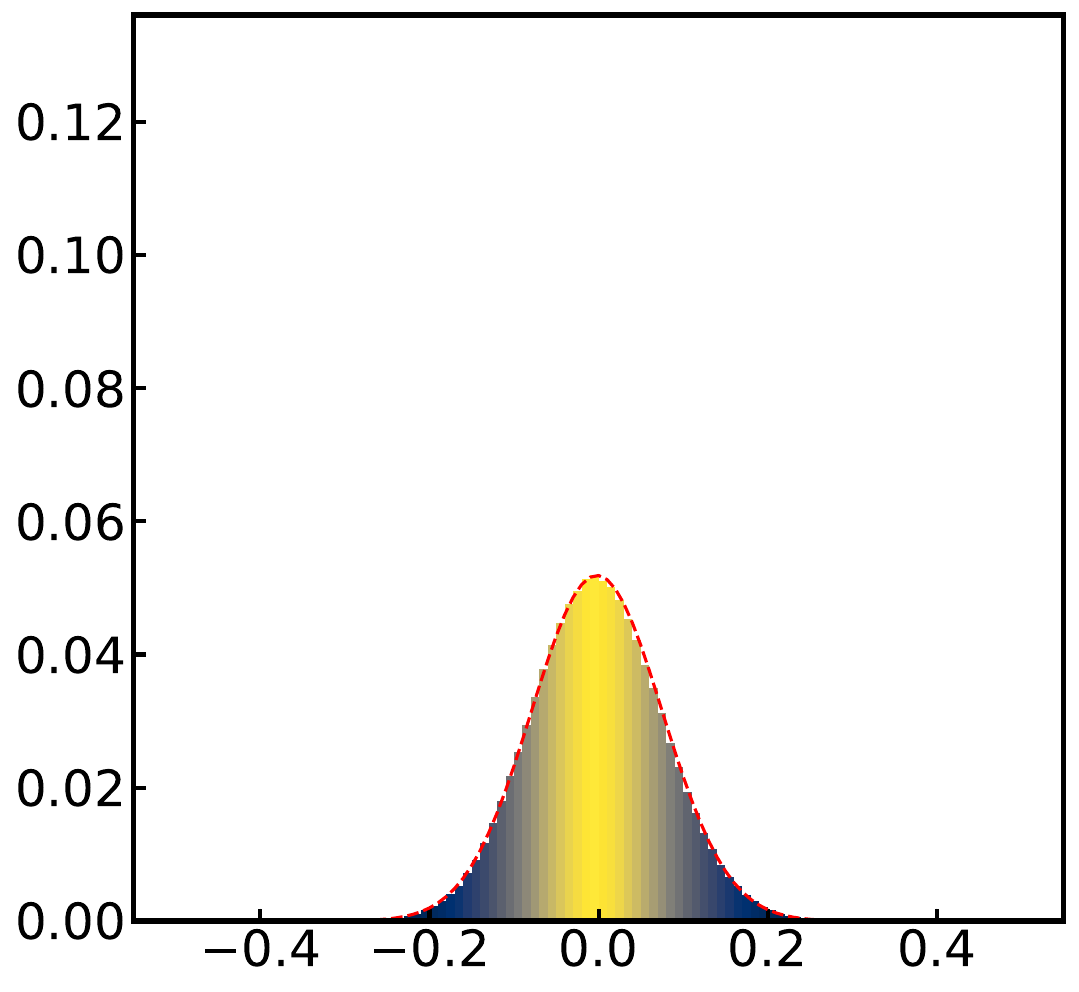}
\put(-37., 90){1200 K}
\put(-62., 80){$\mu$ = -0.0021}
\put(-62., 70){$\sigma$ = 0.077}
\end{subfigure}\hfill
\begin{subfigure}[t]{.22\textwidth}
\centering
\includegraphics[width=\linewidth,trim={2.1cm 1.1cm 0cm 0},clip]{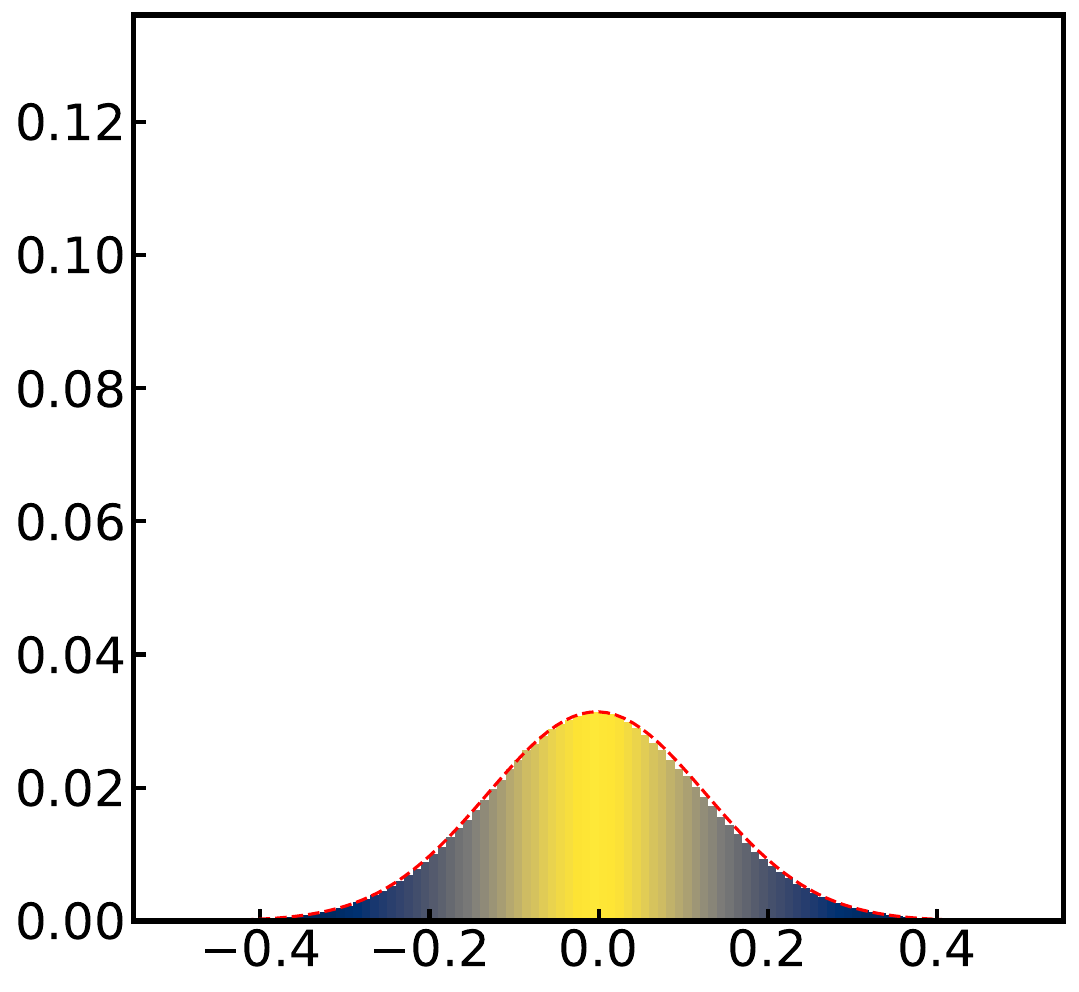}
\put(-37., 90){1200 K}
\put(-62., 80){$\mu$ = -0.0023}
\put(-62., 70){$\sigma$ = 0.13}
\end{subfigure}\hfill
\end{adjustbox}
\begin{adjustbox}{width=0.82\textwidth,center}
\begin{subfigure}[t]{.22\textwidth}
\centering
\includegraphics[width=\linewidth,trim={2.1cm 1.1cm 0cm 0},clip]{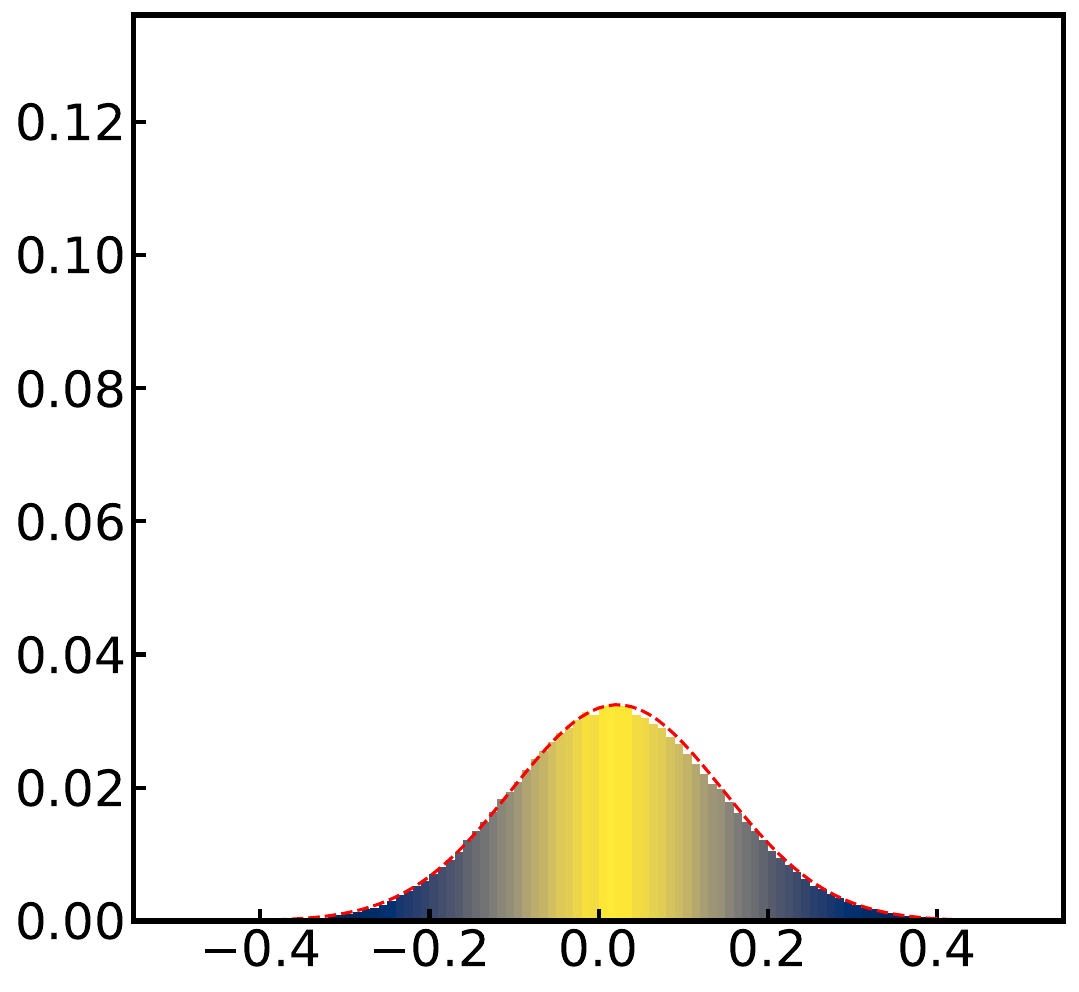}
\put(-37., 90){1500 K}
\put(-62., 80){$\mu$ = 0.022}
\put(-62., 70){$\sigma$ = 0.12}
\put(-128, -1.5){0.00}
\put(-128, 13.5){0.02}
\put(-128, 29){0.04}
\put(-128, 44.5){0.06}
\put(-128, 60){0.08}
\put(-128, 75.5){0.10}
\put(-128, 91){0.12}
\end{subfigure}\hfill
\begin{subfigure}[t]{.22\textwidth}
\centering
\includegraphics[width=\linewidth,trim={2.1cm 1.1cm 0cm 0},clip]{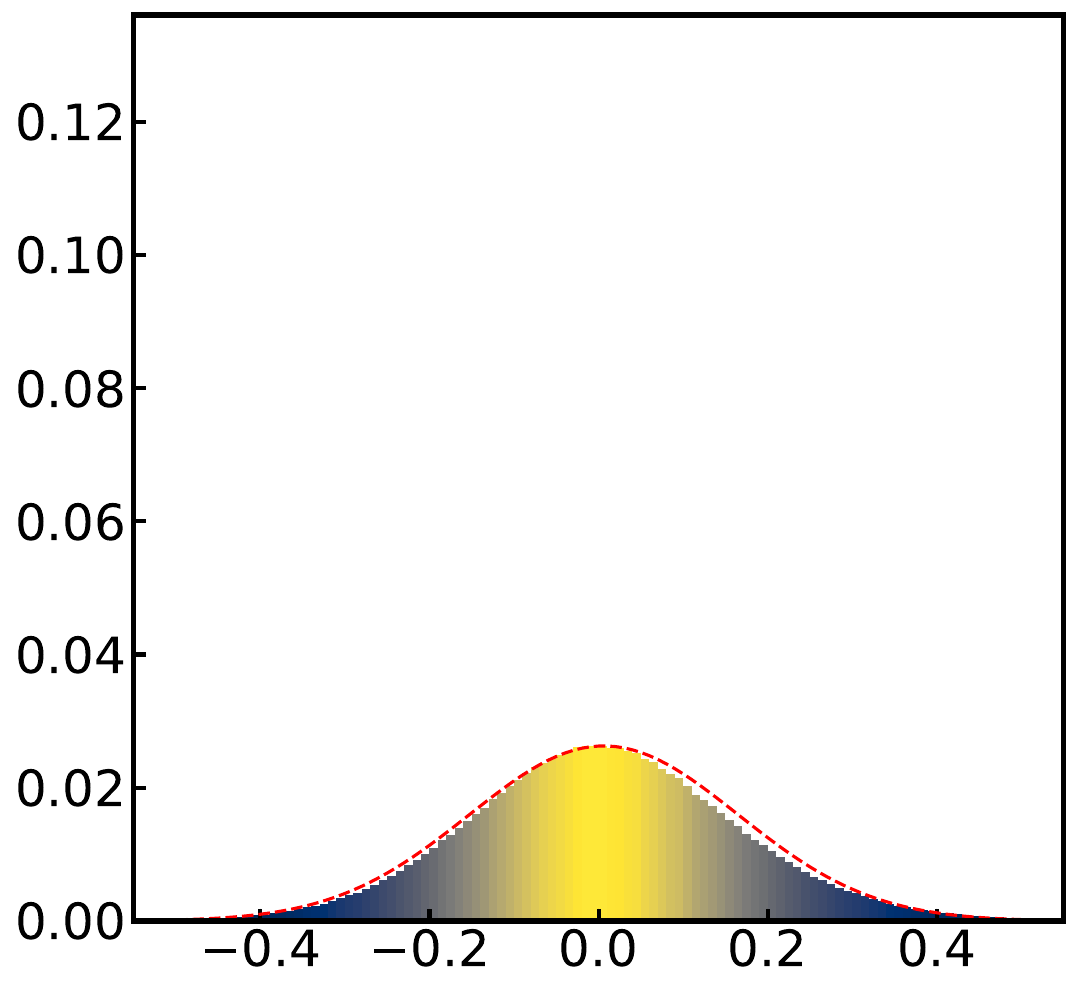}
\put(-37., 90){1500 K}
\put(-62., 80){$\mu$ = 0.0061}
\put(-62., 70){$\sigma$ = 0.16}
\end{subfigure}\hfill
\begin{subfigure}[t]{.22\textwidth}
\centering
\includegraphics[width=\linewidth,trim={2.1cm 1.1cm 0cm 0},clip]{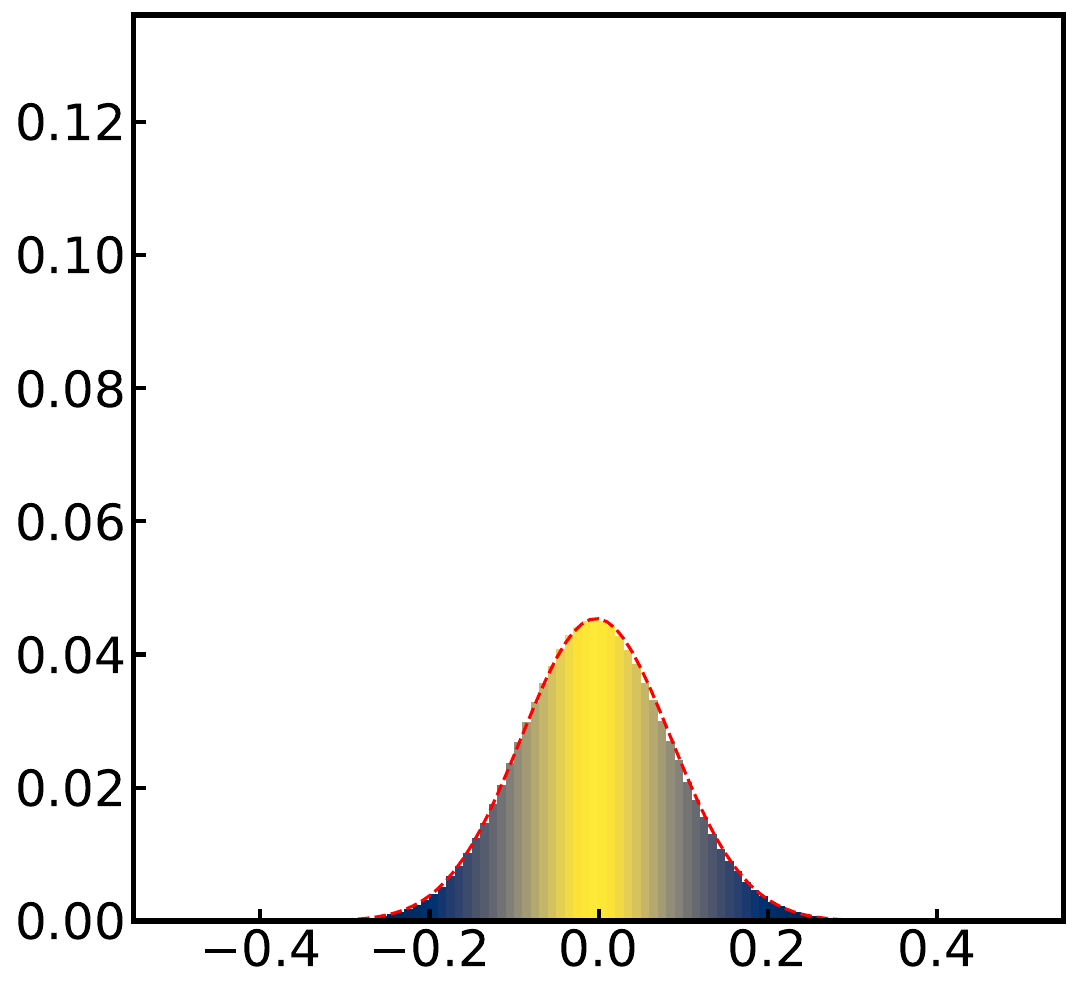}
\put(-37., 90){1500 K}
\put(-62., 80){$\mu$ = -0.003}
\put(-62., 70){$\sigma$ = 0.088}
\end{subfigure}\hfill
\begin{subfigure}[t]{.22\textwidth}
\centering
\includegraphics[width=\linewidth,trim={2.1cm 1.1cm 0cm 0},clip]{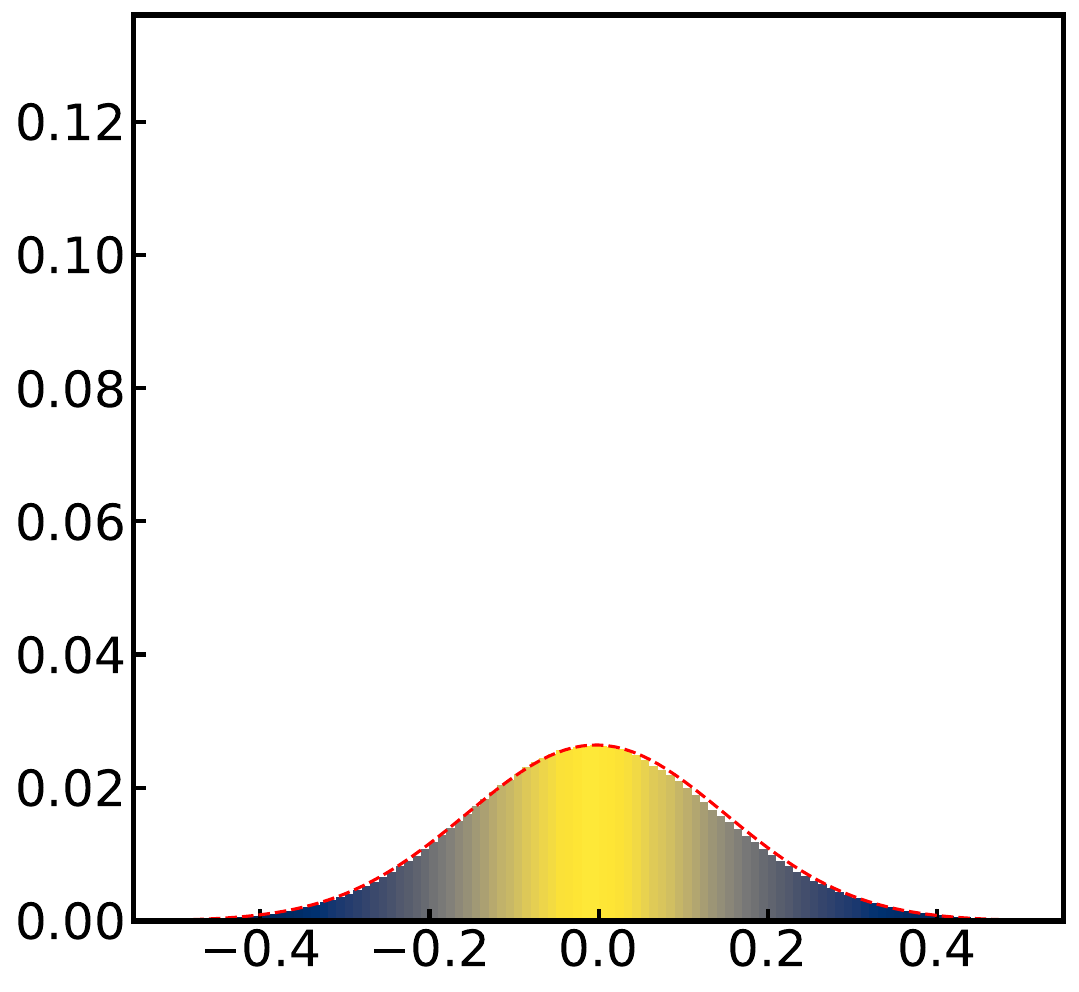}
\put(-37., 90){1500 K}
\put(-62., 80){$\mu$ = -0.0032}
\put(-62., 70){$\sigma$ = 0.15}
\end{subfigure}\hfill
\end{adjustbox}
\begin{adjustbox}{width=0.82\textwidth,center}
\begin{subfigure}[t]{.22\textwidth}
\centering
\includegraphics[width=\linewidth,trim={2.1cm 1.1cm 0cm 0},clip]{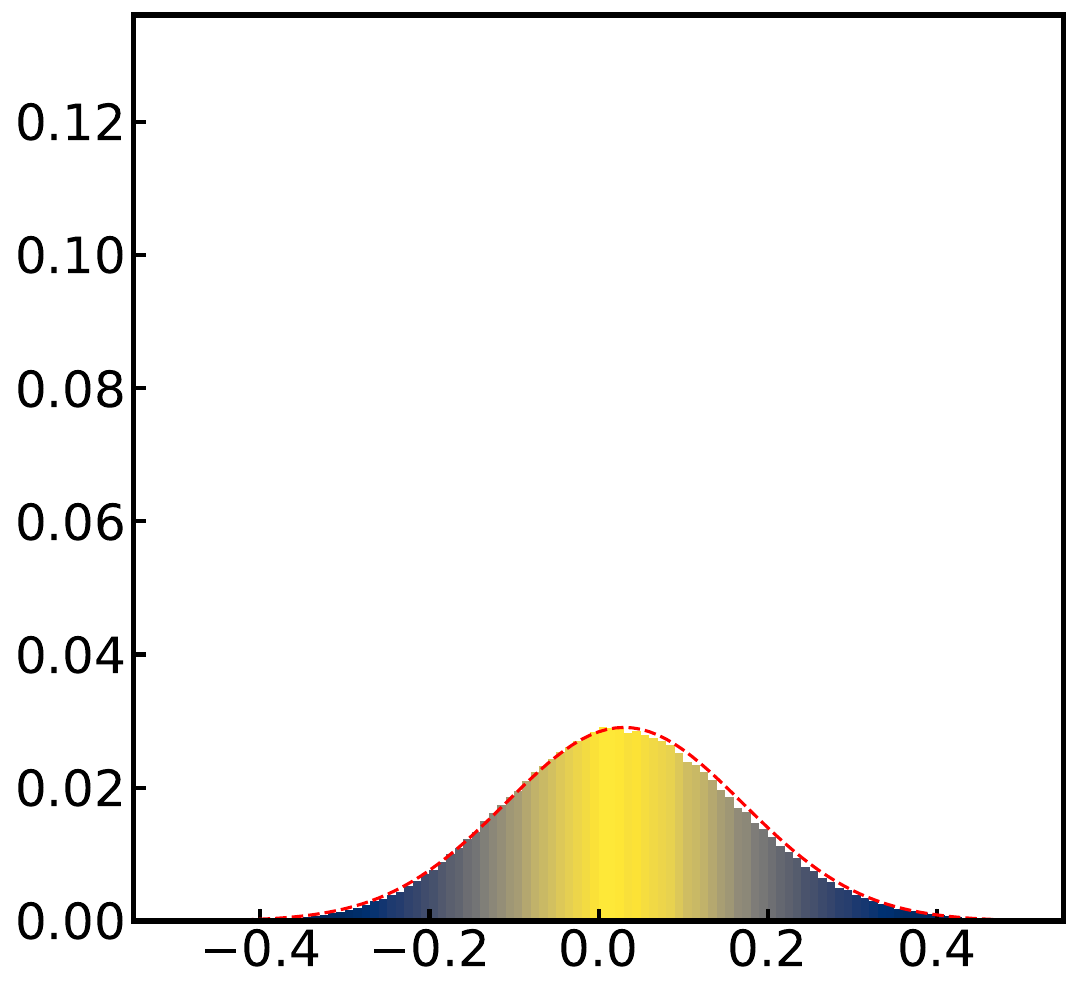}
\put(-37., 90){1800 K}
\put(-62., 80){$\mu$ = 0.03}
\put(-62., 70){$\sigma$ = 0.14}
\put(-104, -7){-0.4}
\put(-84, -7){-0.2}
\put(-62, -7){0.0}
\put(-42, -7){0.2}
\put(-22, -7){0.4}
\put(-128, -1.5){0.00}
\put(-128, 13.5){0.02}
\put(-128, 29){0.04}
\put(-128, 44.5){0.06}
\put(-128, 60){0.08}
\put(-128, 75.5){0.10}
\put(-128, 91){0.12}

\caption{
$\bm v_2(a)\bm v_{1}(a')$
}
\end{subfigure}\hfill
\begin{subfigure}[t]{.22\textwidth}
\centering
\includegraphics[width=\linewidth,trim={2.1cm 1.1cm 0cm 0},clip]{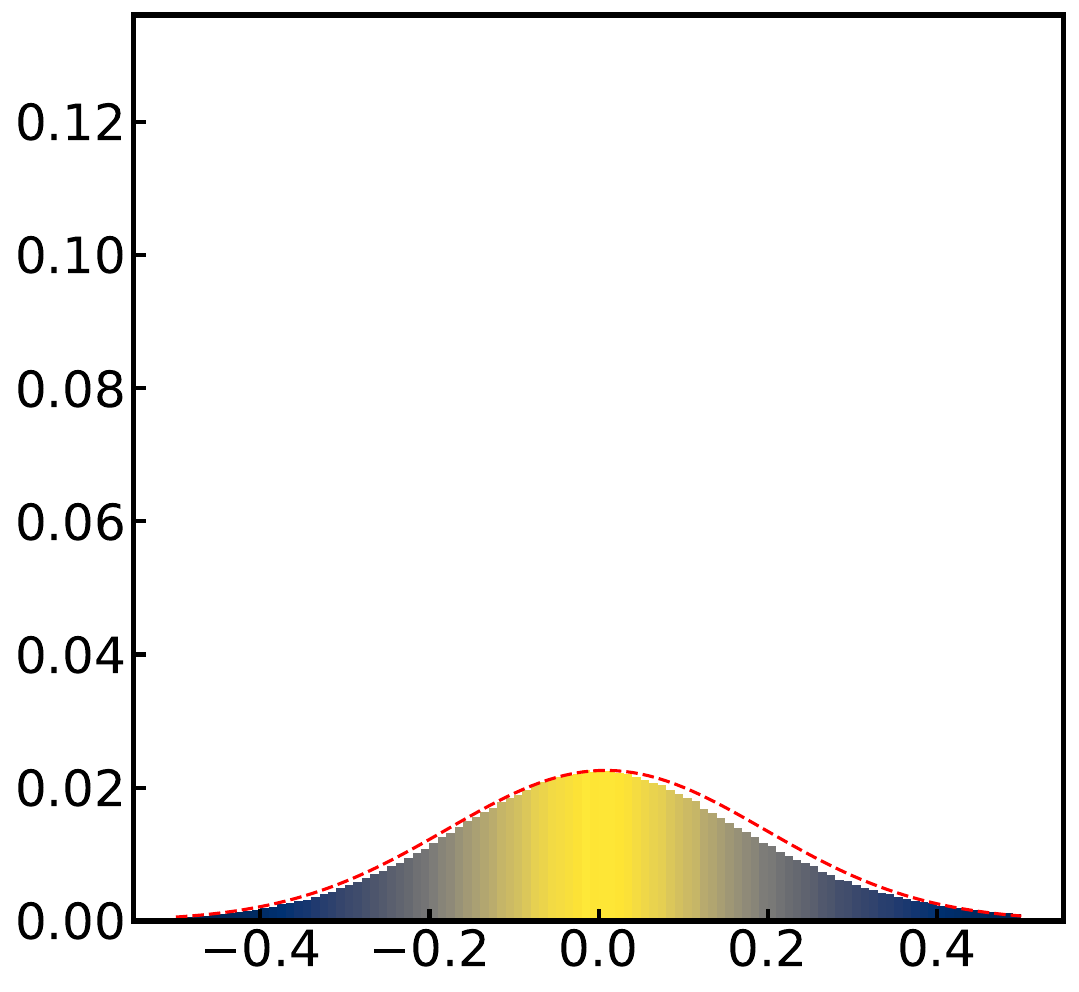}
\put(-37., 90){1800 K}
\put(-62., 80){$\mu$ = 0.0084}
\put(-62., 70){$\sigma$ = 0.19}
\put(-104, -7){-0.4}
\put(-84, -7){-0.2}
\put(-62, -7){0.0}
\put(-42, -7){0.2}
\put(-22, -7){0.4}
\caption{
$\bm v_1(a)\bm v_{1}(a')$
}
\end{subfigure}\hfill
\begin{subfigure}[t]{.22\textwidth}
\centering
\includegraphics[width=\linewidth,trim={2.1cm 1.1cm 0cm 0},clip]{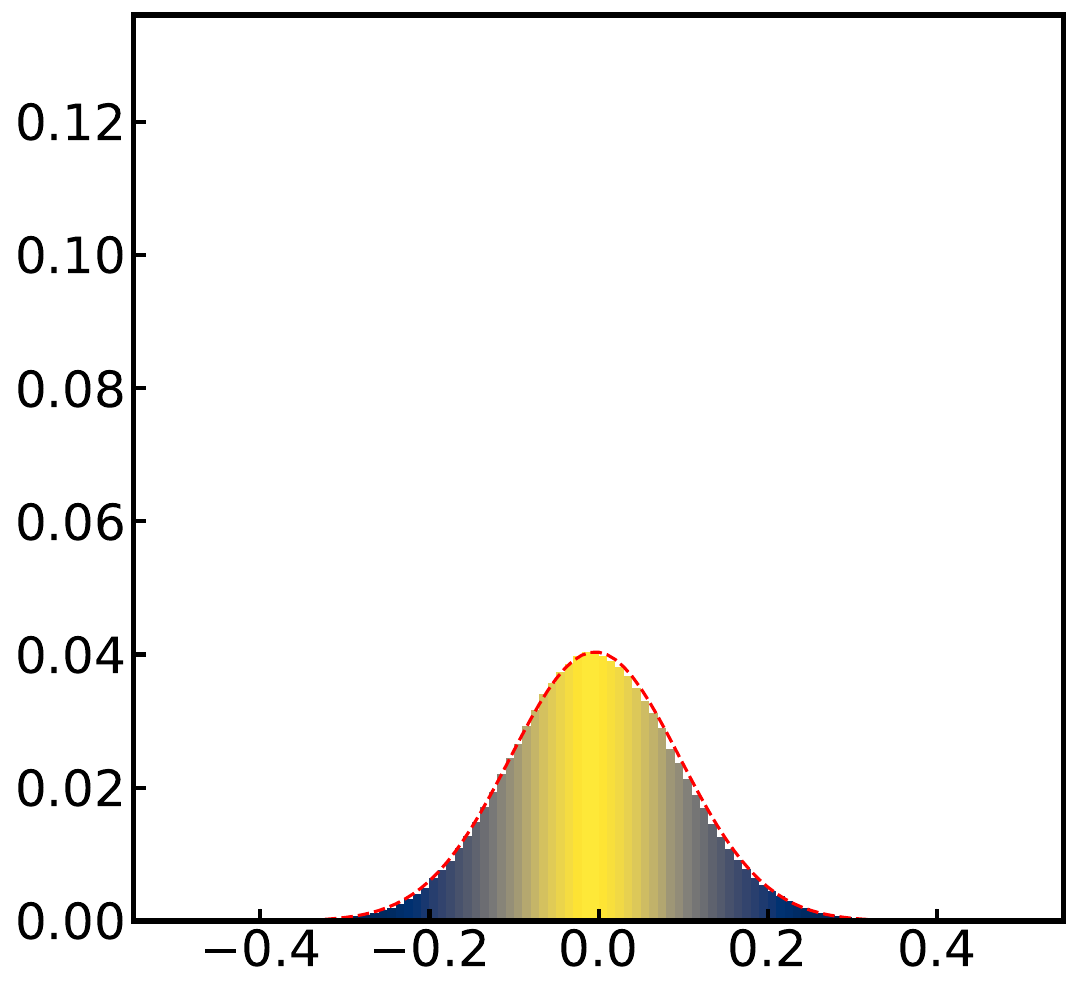}
\put(-37., 90){1800 K}
\put(-62., 80){$\mu$ = -0.0042}
\put(-62., 70){$\sigma$ = 0.1}
\put(-104, -7){-0.4}
\put(-84, -7){-0.2}
\put(-62, -7){0.0}
\put(-42, -7){0.2}
\put(-22, -7){0.4}
\caption{
$\bm v_1(a)\bm v_{3}(a')$
}
\end{subfigure}\hfill
\begin{subfigure}[t]{.22\textwidth}
\centering
\includegraphics[width=\linewidth,trim={2.1cm 1.1cm 0cm 0},clip]{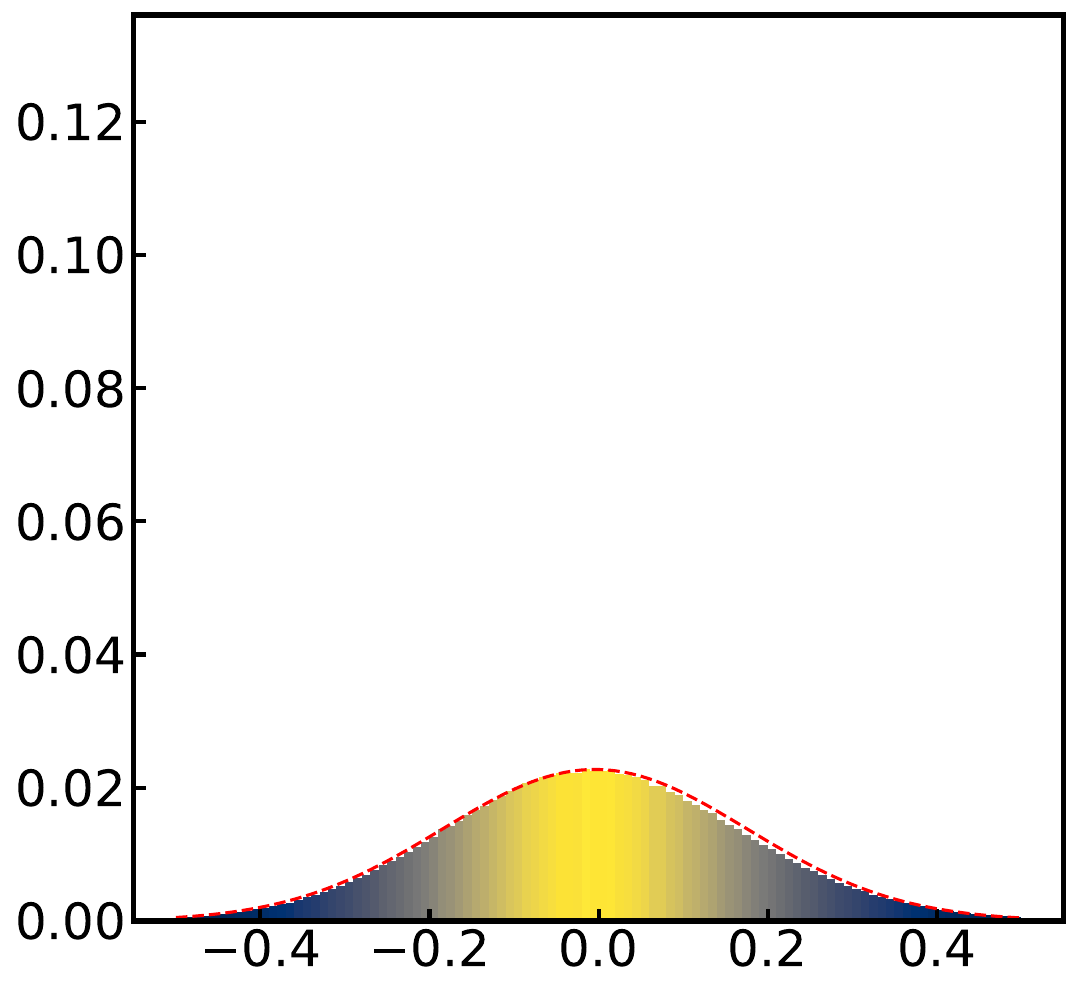}
\put(-37., 90){1800 K}
\put(-62., 80){$\mu$ = -0.0044}
\put(-62., 70){$\sigma$ = 0.18}
\put(-104, -7){-0.4}
\put(-84, -7){-0.2}
\put(-62, -7){0.0}
\put(-42, -7){0.2}
\put(-22, -7){0.4}
\caption{
$\bm v_1(a)\bm v_{2}(a')$
}
\end{subfigure}\hfill
\end{adjustbox}
\caption{The distribution of normal vectors correlations $\bm v_n(a)\bm v_{n'}(a') - \EM[\bm v_n(a) ]\cdot \EM[\bm v_{n'}(a')] $ for equivalent pairs of normals $n, n'$ belonging to nearest-neighboring Voronoi cells $a$ and $a'$, respectively. The vector and atom notation follows those of Fig.~\ref{Fig:SiOverview}(b). The Voronoi cells for all investigated temperatures are calculated using second-generation Car-Parrinello FPMD for a system of N=1000 silicon atoms. The red curve shows the Gaussian fit to the data. }
\label{Fig:normals_distr}
\end{figure*}

\begin{figure}
\centering
\includegraphics[width=.8\linewidth, trim={0cm 0cm 0cm 0cm}, clip]{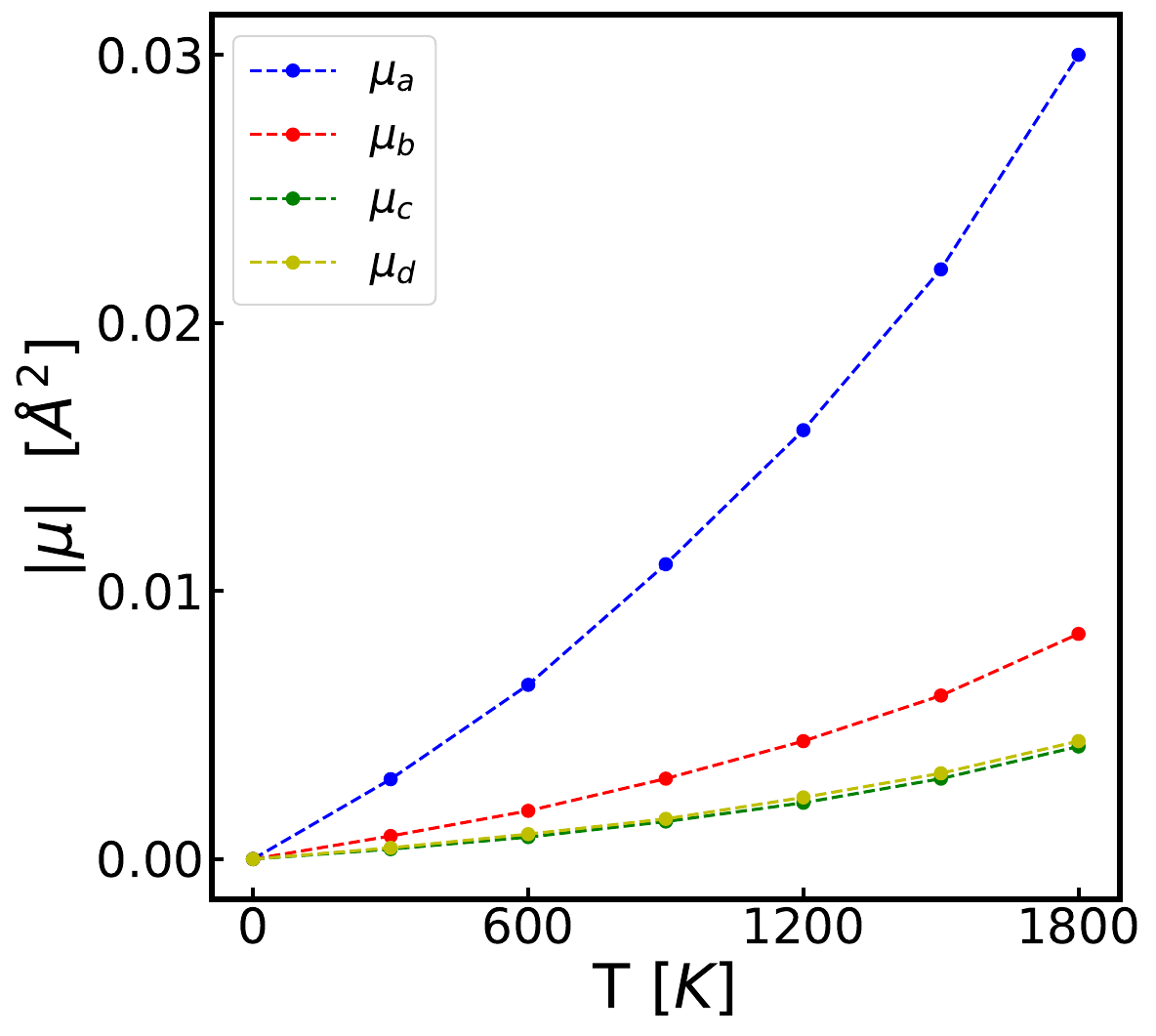}
\caption{The moments of the Gibbs measure as functions of temperature, derived as the absolute values of the means for distributions of Fig.(\ref{Fig:normals_distr}).}
\label{Fig:SigmVsT}
\end{figure}

\section*{Quantifying the Gibbs measure of crystalline silicon} In the case of crystalline silicon, our main conclusion so far is that the configuration space $\Cc$ of the thermally disordered crystal can be encoded in the collection $\{\bm v_n(a)\}$ of the four vectors carried by each atom ``a'' of the crystal. They are displayed in Fig.~\ref{Fig:SiOverview}(b) for two nearest-neighboring atoms, and we will use the same labels to carry on with our discussion.

The data $\{\bm v_n(a)\}$ is over-complete because there are obvious dependencies, such as $\bm v_2(a)+ \bm v_1(a')=\bm 0$ in Fig.~\ref{Fig:SiOverview}(b). As such, the Gibbs measure is only supported by a sub-manifold of $\Cc$, which we now describe.\footnote{We stress that the configuration space $\Cc$ cannot be reduced, because those mentioned dependencies are crucial for the crystal's reconstruction.} A useful observation is that the two atoms seen in Fig.~\ref{Fig:SiOverview}(b) are the pair of atoms contained by the primitive cell shown in Fig.~\ref{Fig:SiOverview}(a), and the large facets of the Voronoi cells, excepting the common ones, coincide with the facets of the repeating cell. To fully specify the positions of any such pairs of atoms, it is enough to retain any four of the eight $\bm v_n$ vectors carried by a pair. We can make this choice for one primitive cell and then use the crystal's translations to select the $\bm v_n$'s for all the other cells. To be concrete, we choose these four vectors to be $\bm v_n(a)$, $n=1,\ldots,4$, in Fig.~\ref{Fig:SiOverview}(b). Since there are no constraints left among the vectors selected by the above procedure, we reached our main statement:

{\bf Statement 3:} {\it The Gibbs measure is singular and supported by a sub-manifold of the configuration space. When restricted to this manifold, the measure is no longer singular and it has a density
\begin{equation}
    {\rm d}{\mathbb P}(\{\bm v_n(p)\}) = \rho(\{\bm v_n(p)\}) \prod_{p,n} {\rm d}^3 \bm v_n(p) ,
\end{equation}
where the product runs over the primitive cells $p$ and the four vectors mentioned above.
}

Our remaining task is to quantify the density of the Gibbs measure, and we will do so by extracting its moments from our FPMD simulations. I.e., we will map the expected values
\begin{equation}
    \EM[v_n(a)_\alpha  v_{n'}(a')_\beta]  = \frac{1}{T}\int_0^T v_n(a;t)_\alpha  v_{n'}(a';t)_\beta \, dt
\end{equation} 
involving the components of the $\bm v$ vectors corresponding to pairs of atoms $(a,a')$, where $a$ and $a'$ can coincide. All these 9 second moments can be conveniently arranged into the $3\times 3$ matrix 
\begin{equation}\label{Eq:2Moments}
    \hat \mu(a,a') := \EM[\bm v(a) \otimes \bm v(a')^T] - \EM[\bm v(a)] \otimes \EM[\bm v(a')^T],
\end{equation}
where the quantities are referenced from their mean values. Since our FPMD simulations assume constant volume, these means are temperature independent and can be read off from the crystal configuration at 0K. We recall that, for a $d$-variate normal distribution with mean $\bar {\bm x}$ and covariance $\hat \Gamma$, i.e.
\begin{equation}\label{Eq:Den1}
    \rho(\bm x)= \frac{(2\pi)^{-\frac{d}{2}}}{|{\rm det}\, \hat \Gamma|^\frac{1}{2}}\exp\big [-\tfrac{1}{2}(\bm x-\bm \mu) ^T \hat \Gamma^{-1} (\bm x - \bm \mu)\big],
\end{equation}
we have 
\begin{equation}
    \EM[\bm x]=\bar{\bm x}, \quad \EM[\bm x \otimes \bm x^T]= \bar{\bm x} \otimes \bar{\bm x}^T + \hat \Gamma.
\end{equation}
Thus, the values we obtain from \eqref{Eq:2Moments} will enable us to directly extract entries of the Gibbs' covariance matrix, if our distribution is well approximated by a multivariate normal distribution.

Using symmetry considerations, we can drastically reduce the amount of data we need to report. Indeed, \eqref{Eq:2Moments} is invariant to crystal re-orientations by 3D-rotations from the silicon crystal's point-group $O_h^7$. As such, the matrix $\hat \mu(a,a')$ of moments must commute with all those 3D-rotation matrices and, since the fundamental representation of $O_h^7$ is irreducible, this is possible if and only if $\hat \mu(a,a')=\mu(a,a') \, I_{3 \times 3}$, with 
\begin{equation}\label{Eq:Sigma}
    \mu(a,n;a',n') = \EM[\bm v_n(a) \cdot \bm v_{n'}(a')]-\EM[\bm v_n(a) ]\cdot \EM[\bm v_{n'}(a')].
\end{equation}
Furthermore, the moments $\mu(a,n;a',n')$ are identical for pairs of vectors that are related by crystal symmetries. Based on this observation and using the notation of Fig.~\ref{Fig:SiOverview}(b), it is enough to report (a) $\mu_a=\mu(a,1;a,1)$, (b) $\mu_b=\mu(a,1;a',1)$, (c) $\mu_c=\mu(a,1;a',3)$ and (d) $\mu_d=\mu_d(a,1;a',2)$, which encode the fluctuations related to first (case a) and second (cases b-d) nearest-neighboring pairs of atoms. The histograms of the quantities appearing in \eqref{Eq:Sigma}, corresponding to the cases (a-d) we just mentioned, are reported in panels (a-d) of Fig.~\ref{Fig:normals_distr}, respectively. The averages of these distributions supply the moments $\mu$ from \eqref{Eq:Sigma} for the cases (a-d), which coincide with the moments of the Gibbs measure pertaining to pairs of first (case a) and second nearest-neighboring atoms (cases b-d). These values are marked in the graphs of Fig.~\ref{Fig:normals_distr} and are plotted in Fig.~\ref{Fig:SigmVsT} as functions of temperature.

There are two consequential observations. First, the distributions in Fig.~\ref{Fig:normals_distr} are extremely well approximated by Gaussian distributions, confirming again our original finding in \cite{KP2018} that the Gibbs measure corresponds to a multivariate normal distribution. Hence, we do not need to map higher moments. Secondly, the moments corresponding to pairs of second nearest-neighboring atoms are one order of magnitude smaller than the moment corresponding to pairs of first nearest-neighboring atoms. This is to say that, for practical purposes, we do not need to consider higher order pairs of neighboring atoms.

\section*{Concluding remarks and outlook}

The moments of the Gibbs measure we provided can be used to generate realistic thermalized atomic configurations. Indeed, Monte Carlo type algorithms can be used to generate various correlated disordered atomic configurations. Thus, these atomic configurations can be used as the input for static and time-dependent first-principles electronic structure simulations to compute realistic electronic properties. For example, the strong temperature dependence of the moments seen in Fig.~\ref{Fig:SigmVsT} leaves no doubt that the electronic properties computed at 0K, as it is often done in the literature, will have limited relevance at finite temperatures where real devices operate.

In our opinion, it will be interesting and important to test our conjecture about the real-space markers for crystalline phases on a large dataset of compounds in the future. 

Lastly, it has not escaped our attention that the present analysis immediately suggests the usage of Voronoi cell normals as descriptors to represent chemical environments within novel machine learning techniques to compute the potential energy surface and molecular properties of atomistic systems \cite{bartok2017machine, keith2021combining}.

\matmethods{All density functional theory-based FPMD simulations were conducted using the second-generation Car-Parrinello method of K\"uhne and coworkers \cite{CPMD2007, kuhne2014second}, as implemented in the CP2K/\textsc{Quickstep} code \cite{CP2K2020}. Therein, the Kohn-Sham orbitals are expanded in a contracted double-$\zeta$ Gaussian basis with one additional set of polarization functions \cite{vandevondele2007gaussian}, whereas the electron density is represented by plane waves up to a density cutoff of 100~Ry. The local density approximation to the exchange and correlation potential was employed, and the interactions between the valence electrons and the ionic cores substituted by norm-conserving Goedecker-Teter-Hutter-type pseudopotentials \cite{GTH1996}. 

The eventual FPMD simulations were conducted in the canonical ensemble using a modified Langevin equation with a discretized integration timestep of 1.0~fs. Specifically, a cubic cell consisting of 1000 silicon atoms subject to periodic boundary conditions was considered. Using these settings we have performed in total 10 FPMD simulations, each 1.25~ns long, whose temperatures were ranging from its semiconducting crystalline state at 300~K up to its metallic liquid phase at 3000~K. 

The 3D Voronoi tesselation is performed using the Voro++ library \cite{rycroft2009voro++, lu2023extension}, which permits the computation of Voronoi cells and all their corresponding normal vectors for each particle individually. 

}

\showmatmethods{} 

\acknow{Emil Prodan acknowledges financial support from the U.S. National Science Foundation through the grant CMMI-2131760 and from the U.S. Army Research Office through contract W911NF-23-1-0127. The authors are grateful for the generous allocation of computing time provided by the Paderborn Center for Parallel Computing (PC2).}

\showacknow{} 

\bibliography{pnas-sample}

\begin{thebibliography}{10}

\bibitem{car1985unified}
R Car, M Parrinello, Unified approach for molecular dynamics and
  density-functional theory.
\newblock {\em\protect\JournalTitle{Physical review letters}} \textbf{55}, 2471
  (1985).

\bibitem{sugino1995ab}
O Sugino, R Car, Ab initio molecular dynamics study of first-order phase
  transitions: melting of silicon.
\newblock {\em\protect\JournalTitle{Physical review letters}} \textbf{74}, 1823
  (1995).

\bibitem{KP2018}
TD K\"uhne, E Prodan, Disordered crystals from first principles i: Quantifying
  the configuration space.
\newblock {\em\protect\JournalTitle{Annals of Physics}} \textbf{391}, 120--149
  (2018).

\bibitem{KHP2020}
TD K\"uhne, J Heske, E Prodan, Disordered crystals from first principles ii:
  Transport coefficients.
\newblock {\em\protect\JournalTitle{Annals of Physics}} \textbf{421}, 168290
  (2020).

\bibitem{Bellissard2015}
JV Bellissard, {\em Delone Sets and Material Science: a Program}, eds.{} J
  Kellendonk, D Lenz, J Savinien.
\newblock (Springer Basel, Basel), pp. 405--428 (2015).

\bibitem{Barnsely1993}
MF Barnsely, {\em Fractals everywhere}.
\newblock (Academic Press, London), (1993).

\bibitem{Lenz2003}
D Lenz, P Stollmann, Delone dynamical systems and associated random operators
  in {\em Operator Algebras and Mathematical Physics}.
\newblock (Theta, Bucharest), pp. 267--285 (2003).

\bibitem{FittingJOM1999}
WD D.W.~Fitting, T Siewert, Monitoring the solidification of single-crystal
  castings using high-energy x-ray diffraction.
\newblock {\em\protect\JournalTitle{JOM}} \textbf{51} (1999).

\bibitem{Singh93}
J Singh, {\em Physics of semiconductors and their heterostructures}.
\newblock (McGraw-Hill, New York), (1993).

\bibitem{wondratschek2004international}
H Wondratschek, U M{\"u}ller, U internationale~de cristallographie, {\em
  International tables for crystallography}.
\newblock (Wiley Online Library) Vol.{}~1, (2004).

\bibitem{CPMD2007}
TD K\"uhne, M Krack, FR Mohamed, M Parrinello, Efficient and accurate
  car-parrinello-like approach to born-oppenheimer molecular dynamics.
\newblock {\em\protect\JournalTitle{Phys. Rev. Lett.}} \textbf{98}, 066401
  (2007).

\bibitem{kuhne2014second}
TD K{\"u}hne, Second generation car--parrinello molecular dynamics.
\newblock {\em\protect\JournalTitle{Wiley Interdisciplinary Reviews:
  Computational Molecular Science}} \textbf{4}, 391--406 (2014).

\bibitem{bellissard2017anankeontheoryviscosityliquids}
JV Bellissard, Anankeon theory and viscosity of liquids: a toy model (2017).

\bibitem{bartok2017machine}
AP Bart{\'o}k, et~al., Machine learning unifies the modeling of materials and
  molecules.
\newblock {\em\protect\JournalTitle{Science advances}} \textbf{3}, e1701816
  (2017).

\bibitem{keith2021combining}
JA Keith, et~al., Combining machine learning and computational chemistry for
  predictive insights into chemical systems.
\newblock {\em\protect\JournalTitle{Chemical reviews}} \textbf{121}, 9816--9872
  (2021).

\bibitem{CP2K2020}
TD K\"uhne, et~al., Cp2k: An electronic structure and molecular dynamics
  software package - quickstep: Efficient and accurate electronic structure
  calculations.
\newblock {\em\protect\JournalTitle{J. Chem. Phys.}} \textbf{152}, 194103
  (2020).

\bibitem{vandevondele2007gaussian}
J VandeVondele, J Hutter, Gaussian basis sets for accurate calculations on
  molecular systems in gas and condensed phases.
\newblock {\em\protect\JournalTitle{The Journal of chemical physics}}
  \textbf{127} (2007).

\bibitem{GTH1996}
S Goedecker, M Teter, J Hutter, Separable dual-space gaussian pseudopotentials.
\newblock {\em\protect\JournalTitle{Phys. Rev. B}} \textbf{54}, 1703 (1996).

\bibitem{rycroft2009voro++}
CH Rycroft, Voro++: A three-dimensional voronoi cell library in c++.
\newblock {\em\protect\JournalTitle{Chaos: An interdisciplinary journal of
  nonlinear science}} \textbf{19}, 41111 (2009).

\bibitem{lu2023extension}
J Lu, EA Lazar, CH Rycroft, An extension to voro++ for multithreaded
  computation of voronoi cells.
\newblock {\em\protect\JournalTitle{Computer Physics Communications}}
  \textbf{291}, 108832 (2023).

\end{thebibliography}

\end{document}